\documentclass{aa}
\usepackage{graphicx}
\usepackage{txfonts}
\usepackage{natbib}
\usepackage{longtable}
\usepackage[shortlabels]{enumitem}
\usepackage{siunitx}

\begin{document}

%\usepackage{txfonts}

%\setcitestyle{citesep={,}}

%%%%% AUTHORS - PLACE YOUR OWN PACKAGES HERE %%%%%

% Only include extra packages if you really need them. Common packages are:
        % Including figure files
% Extra maths symbols

%%%%%%%%%%%%%%%%%%%%%%%%%%%%%%%%%%%%%%%%%%%%%%%%%%

\title{Galactic Archaeology with asteroseismic ages: evidence for delayed gas infall in the formation of the Milky Way disc}
\author { E. Spitoni\inst{1}  \thanks {email to: spitoni@phys.au.dk} \and
  V. Silva  Aguirre\inst{1} \and F. Matteucci\inst{2, 3, 4} \and F. Calura
  \inst{5} \and V. Grisoni \inst{2}}
\institute{Stellar Astrophysics Centre, Department of Physics and
  Astronomy, Aarhus University, Ny Munkegade 120, DK-8000 Aarhus C,
  Denmark \and Dipartimento di Fisica, Sezione di Astronomia,  Universit\`a di Trieste, via
  G.B. Tiepolo 11, I-34131, Trieste, Italy \and I.N.A.F. - Osservatorio
  Astronomico di Trieste, via G.B. Tiepolo 11, I-34131, Trieste,
  Italy \and I.N.F.N. -  Sezione di Trieste, Via Valerio 2, I-34100
  Trieste \and I.N.A.F. - Osservatorio Astronomico di Bologna, Via Gobetti 93/3, 40129 Bologna, Italy}

\date{Received xxxx / Accepted xxxx}

\abstract
{Precise stellar ages from asteroseismology have become available and can help setting stronger constraints on the evolution of the Galactic disc components. Recently, asteroseismology has confirmed a clear age difference in the solar annulus between two distinct sequences in the [$\alpha$/Fe] versus
[Fe/H] abundance ratios relation: the high-$\alpha$ and low-$\alpha$ stellar populations.}
{We aim at reproducing these new data with chemical evolution models including different assumptions for the history and number of accretion events.}
{We tested two different approaches: a revised version of the ``two-infall'' model  where the high-$\alpha$ phase forms by a fast gas accretion episode and the low-$\alpha$ sequence follows later from a slower gas infall rate, and the parallel formation scenario where the two disc sequences form coevally and independently.}
{The revised ``two-infall'' model including uncertainties in age and metallicity is capable of reproducing: i) the [$\alpha$/Fe] vs. [Fe/H]  abundance relation at different Galactic epochs, ii) the age$-$metallicity relation and the time evolution [$\alpha$/Fe]; iii) the age distribution of the high-$\alpha$ and low-$\alpha$ stellar populations, iv) the metallicity distribution function. The parallel approach is not capable of properly reproduce the stellar age distribution, in particular at old ages.}
{The best chemical evolution model is the revised ``two-infall'' one, where a consistent delay of $\sim$4.3 Gyr in the beginning of the second gas accretion episode is a crucial assumption to reproduce stellar abundances and ages.}


\titlerunning{Galactic Archaelogy with asteroseismic ages}

\authorrunning{Spitoni et al.}

\maketitle

\section{Introduction}
The main goal of  Galactic Archaeology is to find and interpret  signatures for the formation and evolution of our Galaxy from the observed chemical abundances and kinematics of resolved stellar populations \citep{BlandHawthorn:2016iq}. Tracing the history of the formation and evolution of our Galaxy is a fundamental step to understand the evolution of the Universe.

Each stellar atmosphere carries the enrichment history of the interstellar medium (ISM) from which it was formed. Once a star is born, and although its interior composition evolves, its atmosphere is negligibly polluted by the effects of stellar evolution. For this reason, stars are the surviving relics of formation and accretion episodes, and carry the most genuine signature of the processes which determined the formation and regulated the evolution of the various components of our Galaxy.

Chemical abundances are now routinely measured in stars belonging to the Galactic disc by spectroscopic surveys such as the Apache Point Observatory Galactic Evolution Experiment project \citep[APOGEE][]{Majewski:2017ip}, the Gaia-ESO survey \citep{gilmore2012} and Galactic Archaeology with HERMES \citep[GALAH][]{buder2018}. Combining this wealth of information with kinematic properties of stellar ``fossil'' relics provided by the second Gaia data release \citep[DR2][]{GaiaCollaboration:2018dt} offers an unparalleled opportunity to test galaxy formation models. This synergy has the potential for setting strong constraints to the history of star formation and unravel the importance of the various physical processes that led to the formation of our Galaxy.

 Previous works accomplished the determination of stellar ages based purely on spectroscopic or photometric information but for limited number of stars in the solar vicinity: \citet{bensby2014}  data sample is composed by 714 stars , \citet{bergemann2014} one by 144 stars, and \citet{haywood2013} sample is composed by 363 stars.
Due to the difficulties in determining ages for field stars based purely on spectroscopic or photometric information, most studies of the Milky Way disc have focused on identifying different populations in the solar neighborhood using chemistry and kinematics tagging. Recent data from spectroscopy pointed out the existence of a clear distinction between two sequences of disc stars in the [$\alpha$/Fe] versus [Fe/H]  space,  e.g. the Gaia Eso Survey  \citep{RecioBlanco:2014dd,RojasArriagada:2016eq,RojasArriagada:2017ka}, the APOGEE project \citep{Nidever:2014fj,hayden2015},  the AMBRE project \citep{Mikolaitis:2017gd}. These sequences have been reproduced by tuning different parameters in chemical evolution models \citep[e.g.,][]{Nidever:2014fj,Snaith:2015ki,Haywood:2016el,grisoni2018,grisoni2017}, and recently predicted in the context of cosmological zoom-in simulations of Milky Way type galaxies \citep{grand2018,mackereth2018}.

Different prescriptions can be used in the chemical evolution models to reproduce particular features in the spectroscopic data. For instance, \citet{Snaith:2015ki} and \citet{Haywood:2016el} considered the Galaxy as a closed-box system assuming that the accretion gas episodes are concentrated in the early initial phase of Galactic evolution. Those model are characterized by a significant dip in star formation between the high-$\alpha$ and low-$\alpha$ stars.   On the other hand, several authors have developed models with episodes of exponential infall of gas throughout Galactic
history, i.e. \citet{spitoni2014}, \citet{cote2017}, \citet{rybi2017}, \citet{prantzos2018}. All these models share the common feature of reproducing the observed distribution of stars in chemical space, i.e., the two sequences in the [$\alpha$/Fe] vs [Fe/H] plane. However, they predict different star formation histories and thus different correlations of stellar properties with age.

A step further from the constraints provided by abundance ratios and kinematics of stars comes from the  new dimension provided by the asteroseismology: precise stellar ages. 
Detailed asteroseismic analysis is a powerful tool to probe stellar interiors, since the oscillation frequencies are closely related to physical properties of stars via the density and sound speed profiles \citep[see][and references therein]{Aerts:2010uw,Chaplin:2013gz}. Since these quantities are tightly linked to the stellar mass and evolutionary  stage they can deliver precise ages of stars by comparing their oscillation spectrum with predictions of stellar models \citep[e.g.,][]{Casagrande:2014bd,Serenelli:2017cn,2018arXiv180409983P}. For field red giants, asteroseismic age uncertainties are at the level of $\sim$25\% \citep[e.g.,][]{casagrande2016,2017A&A...597A..30A,2018MNRAS.475.5487S}.

With the recently established synergy of asteroseismic observations and high-resolution spectroscopy surveys, it has become possible to determine stellar properties for thousands of red giants in different regions of the Galaxy. Combining atmospheric parameters from APOGEE with data from the {\it Kepler} satellite, \citet[][hereafter VSA18]{2018MNRAS.475.5487S} found that the two distinct sequences in the [$\alpha$/Fe] vs [Fe/H] abundance ratios plane are characterized by a  clear age difference. The low-$\alpha$ sequence age distribution peaks at $\sim$2 Gyr, whereas the high-$\alpha$ one does it at $\sim$11 Gyr. 
This was the first confirmation using asteroseismology of the age gap between these chemically selected populations, as already pointed out by e.g. \citet{fuhr1998} using a very local ($\sim$ 25pc) but complete sample over in the solar neighbourhood.
This age gap still needs to be confirmed in other Galactic regions, and the advent of asteroseismology as a tool for Galactic archaeology appears as the most promising route to test this paradigm across the Milky Way \citep[e.g.,][]{Miglio:2013hh,Casagrande:2014bd,2015ApJ...809L...3S}.

In this paper we test our chemical evolution models with the aim of reproducing the new data by \citet{2018MNRAS.475.5487S}. We discuss two different approaches  to reproduce the high and low-$\alpha$ sequences: i) the ``two-infall'' approach and the ii) parallel one. In the latter, the various Galactic stellar components begin to form at the same time but evolve in parallel at different rates. On the other hand, the revised ``two-infall''  model by \citet{grisoni2017} follows a sequential scenario: first the thick disc is formed by a gas infall episode, and later on a totally independent gas accretion event creates the thin disc on longer time-scales.

Our paper is organized as follows. In Section~\ref{s:apokasc} we present the APOKASC sample by VSA18. In Section 3 we describe the details of our adopted chemical evolution model for the solar neighbourhood. Section 4 presents our results related to the best ``two-infall'' model, and in Section 5 the model results in which we take into account the observational errors for the age and metallicity estimates are drawn. In Section 6 we present our results varying the infall time-scales of the different accretion episodes.  In Section 7 the ``parallel'' chemical evolution model results are reported. Finally, our conclusions are drawn in Section 8.
\section{The APOKASC sample by Silva Aguirre et al. (2018)}\label{s:apokasc}
\citet{pinso2014} presented the first APOKASC (APOGEE+ {\it Kepler} Asteroseismology Science Consortium) catalogue of spectroscopic and asteroseismic properties of 1916 red giants observed in the {\it Kepler}  field. The updated APOKASC sample presented by VSA18 is composed by 1989 red giants, with stellar properties determined combining the photometric, spectroscopic, and asteroseismic observables in the BAyesian STellar Algorithm \citep[{\tt BASTA},][]{silvaaguirre2015,silvaAguirre2017} framework.

They associated to this sample proper motions from the first DR of Gaia \citep{linde2016,gaia2016} and The Fourth US Naval Observatory CCD Astrograph Catalog (UCAC-4) catalogue \citep{zach2013}. A first pruning procedure was applied by retaining only stars with precise kinematic information available.

To ensure that the chosen sample was representative
of the physical and kinematic characteristics of the true
underlying population of red giants in that direction of the sky,
a
selection function was applied adopting the  \citet{casagrande2016} method. Briefly, they corrected for the selection of oscillating giants with available
APOGEE spectra, and after for the target selection effects of the
{\it Kepler} spacecraft as a function of distance.
The final sample presented by VSA18 is
composed by 1197 stars.  Due to its large number of stars with available ages and correction for selection effects compared to other studies \citep[][]{haywood2013,bergemann2014,bensby2014}, the APOKASC sample is to be regarded as extremely valuable and particularly suited for chemical evolution studies.

\begin{figure}
\begin{centering}
\includegraphics[scale=0.44]{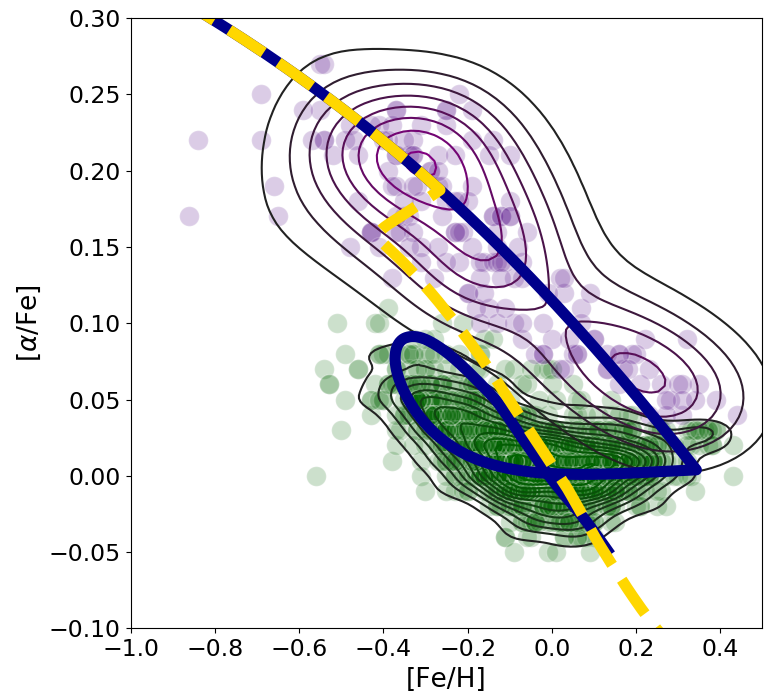}
\caption{The observed [$\alpha$/Fe] vs. [Fe/H] abundance ratios presented by VSA18, compared with our fiducial chemical evolution model  (blue line) and the classical  two-infall model (yellow line) (see Section~\ref{s:revised}).  The purple filled circles are the observed ``high-$\alpha$'' stars,  whereas the green filled circles are the observed ``low-$\alpha$'' stars. 
The  black-purple and black-green contour lines enclose the observed high-$\alpha$ and low-$\alpha$ stars, respectively. }
\label{AFEFE1}
\end{centering}
\end{figure}

In Fig.~\ref{AFEFE1} the observed [$\alpha$/Fe] vs [Fe/H]
abundance ratios for the stars presented by VSA18 are reported.  Here, it is assumed that $\alpha$ abundances are given by the sum of the individual Mg and Si abundances \citep{salaris2018}. The figure shows the disc components selected on the basis of their chemical properties: the high-$\alpha$ stars have ages of $\sim$11 Gyr, while the low-$\alpha$ sequence peaks at $\sim$2 Gyr. For guidance, we also depict the resulting prediction of our fiducial chemical evolution model (see Section~\ref{s:chemmod} for details).

In the sample that we will use throughout this work we have not taken
into account the so called ``young $\alpha$ rich'' (Y$\alpha$R) stars.
Understanding the origin of these stars has
been the subject of a number of recent studies and they have been
attributed to stars migrated  from the Galactic bar \citep{chiappini2015}
as well as evolved blue stragglers \citep{martig2015,chiappini2015,yong2016,jofre2016}. In the former case,
it is believed that these stars formed in reservoirs of almost inert
gas close to the end of the Galactic bar, while the latter scenario
proposes that the Y$\alpha$R stars are the product of mass transfer or
stellar merger events.

In the next Sections we want to present a chemical evolution model
focused only on the solar neighbourhood stars. Our aim is to provide a
simple and valid scenario capable to explain the majority of the stars
and the new stellar ages contraints as  provided by  asteroseismology. 
\section{The chemical evolution model for the solar neighbourhood}\label{s:chemmod}
We concentrate our study on the evolution with time of the solar annulus, defined here by a annular ring 2 kpc wide centered at 8 kpc from the Galactic centre. Our chemical evolution model  is capable of tracing the abundance change with time of several chemical species (H, D, He, Li, C, N, O, $\alpha$-elements, Fe, Fe-peak elements, s- and r- process elements). We take into account detailed nucleosynthesis from low and intermediate mass stars, Type Ia SNe (originating
from white dwarfs in binary systems) and Ib/c and II SNe (originating from core-collapse of massive stars). 
The contribution of Type Ia SNe  was first introduced by \citet{matteucci1986}. Here, the
rate is calculated by assuming the single-degenerate model for the
progenitor of these SNe, namely a  Carbon-Oxygen White Dwarf (C-O WD) plus a red giant companion in
a binary system,  and  is expressed as:
\begin{equation}
R_{SNeIa}=A_{Ia}\int\limits^{M_{BM}}_{M_{Bm}}\phi(M_{B})\left[ \int\limits^{0.5}_{\mu_{m}}f(\mu)\psi(t-\tau_{M_{2}})d\mu \right]
dM_{B},
\end{equation}
where $M_{2}$ is the mass of the secondary, $M_{B}$ is the total mass of the binary
system, $\mu=M_{2}/M_{B}$, $\mu_{m}=max\left[M_{2}(t)/M_{B},(M_{B}-0.5M_{BM})/M_{B}\right]$, 
$M_{Bm}= 3 M_{\odot}$, $M_{BM}= 16 M_{\odot}$. The IMF is represented by $\phi(M_{B})$
and refers to the total mass of the binary system for the computation of the Type Ia SN rate,
$f(\mu)$ is the distribution function for the mass fraction of the secondary:
 $A_{Ia}$ is the fraction of systems in the appropriate mass
range, which can give rise to Type Ia SN events.  Details about the assumed parameter values   can be found in \citet{spitoni2009}.

The star formation rate (SFR) is implemented with \citet{kenni1998} law:
\begin{equation}
\psi(t) =\nu \sigma_{g}^{k},
\label{k1}
\end{equation}
where $\nu$ is the star formation efficiency,  $\sigma_g$ is the surface
gas density, and $k$ is the gas surface exponent with an exponent $k$=1.5. For the IMF we use that of \citet{scalo1986} (constant in time and space).

The temporal evolution of the surface density of a certain chemical element $\Sigma_i(R,t)$ is given by the following expression:
\begin{equation}
    \dot \Sigma_i(R,t)= -X_i(R,t) \times \mbox{SFR}(R,t) +
    \mathcal{R}_i(R,t)  +\mathcal{B}_i(R,t)\,,
\label{CHEM}
\end{equation}
where $X_i(R,t)$  is the abundance by mass of the element $i$,  $\mathcal{R}_i(R,t)$ is the returned fraction, and $\mathcal{B}_i(R,t)$ is the infall rate term.

We assume that the Galaxy is an ``open'' system and forms by gas accretion episodes that follow  an exponentially decreasing infall rate as a function of time. This fundamental assumption adopted in most of the detailed numerical
chemical evolution models of our Galaxy \citep{chiappini1997,romano2010,spitoni2014,grisoni2017}  helps solving the G dwarf distribution problem. Moreover,  \citet{colavitti2008} showed  that the ``two-infall'' model of \citet{chiappini1997} is qualitatively in agreement with results of the GADGET2 \citep{springel2005} cosmological hydrodynamical simulations when the standard cosmological parameters from WMAP \citep{spergel2007} are assumed.
Although  an important ingredient of the \citet{Nidever:2014fj} chemical evolution model to reproduce the APOGEE data was the inclusion of  Galactic winds proportional to the SFR coupled to a variable loading factor, in this paper we do not consider outflows. In fact, \citet{melioli2008,melioli2009} and \citet{spitoni2008,spitoni2009}, studying the Galactic fountains processes originated  by the explosions of Type II SNe in OB associations, found that the ejected metals fall back close to the same Galactocentric region where they are delivered and thus not modify significantly  the chemical evolution of the disc as a whole. Therefore, we do not take into account events of gas outflows in
our models.
\subsection{An updated ``two-infall'' model for the high-$\alpha$ and low-$\alpha$ components formation}
We present here the ``two-infall'' chemical evolution model
designed  to reproduce the high-$\alpha$ and  low-$\alpha$ stars sequences presented by VSA18. The infall rate $\mathcal{B}_i(t,i)$, to be inserted in the right side of eq.(\ref{CHEM}) is:
\begin{equation}
\mathcal{B}_i(t,i)=(X_i)_{inf} \left[ c_1 \, e^{-t/ \tau_{D1}}+ c_2 \, e^{-(t-t_{max})/ \tau_{D2}} \right],
\label{a}
\end{equation}
where $(X_i)_{inf}$ is the abundance by mass of the element $i$ of the infall gas that here  is assumed to be primordial, while $t_{max}$ is the time for the maximum infall on the  second accretion episode, i.e. indicates the delay of 
the beginning of the second infall. The typical value assumed for $t_{max}$ in previous "two-infall" models without constraints from stellar ages is $\sim$ 1 Gyr \citep{chiappini2001,romano2010,spitoni2009}.

The quantity $\tau_{D1}$ is the time-scale for the creation of the high-$\alpha$ stars and $\tau_{D2}$ is the timescale for the formation of the low-$\alpha$ disc phase. Finally, the coefficients $c_1$ and $c_2$ are obtained by imposing a fit to the observed current total surface mass density in the solar neighbourhood, adopting the following relations:
\begin{equation}
c_1 =\frac{\Sigma_{tot1}(t_G)}{\tau_{D1} (1- e^{-t_G/\tau_{D1}})},
\label{c1}
\end{equation}
\begin{equation}
c_2 =\frac{\Sigma_{tot2}(t_G)}{\tau_{D2} (1-
  e^{-(t_G-t_{max})/\tau_{D2}})},
\label{c2}
\end{equation}
where $\Sigma_{tot1}(t_G)$ and $\Sigma_{tot2}(t_G)$ are the present
day total surface mass density of the high-$\alpha$ and low-$\alpha$ phases,
respectively.

In this particular model, differently from \citet{Nidever:2014fj}, we do not consider Galactic winds, in fact in presence of winds eqs. (\ref{c1}) and (\ref{c2}) need to be revised.

\subsection{The ``parallel'' formation scenario}

With the aim  of reproducing the data from the AMBRE project in the solar neighbourhood,  \citet{grisoni2017}  tested the possibility  of abandoning a sequential scenario in favor of a picture in which  thick disc and  thin disc stars are described by two  coeval and independent evolutionary phases. This scenario was suggested by the fact that AMBRE data seemed to form two parallel sequences in the [Mg/Fe] vs [Fe/H] abundance ratio relation. In this paper we consider this scenario in the light of the new observational data by VSA18 comparing the stellar ages predicted by the model with the one provided by the asteroseismology.

Therefore we have to solve in this case 2 independent  sets of integro-differential equations presented by eq. (\ref{CHEM}) assuming two distinct infall episodes. The gas infall rates for the high-$\alpha$ and low-$\alpha$ sequences are, respectively:
\begin{equation} \label{par1}
\mathcal{B}_i^{T}(t,i)=(X_i)_{inf} \,  c_{T} \,  e^{-\frac{t}{\tau_T}},
\end{equation}
\begin{equation} \label{par2}
\mathcal{B}_i^{D}(t,i)=(X_i)_{inf} \,  c_{D}\,  e^{-\frac{t}{\tau_D}}.
\end{equation}
The quantities $c_{T}$ and $c_{D}$ along with the parameters $\tau_T$ and $\tau_D$ have the same meaning of the parameters introduced in  eq (\ref{a}). The novelty
introduced by \citet{grisoni2017} with this scenario is the fact that the infall rates of the phases are totally disentangled and coeval. 
\subsection{Nucleosynthesis prescriptions}
For the nucleosynthesis prescriptions of  Fe, Mg and Si we adopted those suggested by \citet{francois2004}, who found that the  yields  at solar metallicity of Type II SNe  of \citet{WW1995} provide the best fit to the data (details related to the adopted  observational data  are in \citet{francois2004}). The authors artificially increased the yields from massive stars of Mg from \citet{WW1995} to reproduce the solar Mg abundance.
 Mg yields from stars in the range 11-20 M$_ {\odot}$ have been increased by a factor of  7  whereas those from stars larger than 20 M$_ {\odot}$  are lower than predicted by \citet{WW1995} (a factor of 2 on average).   No modifications are required for the yields of  Fe, as computed for solar chemical composition.  Concerning Si, only the yields of the very massive stars (M $>$ 40 M$_ {\odot}$) should be increased by a factor of 2. The complete grid of the modified Mg, Si, and Fe yields  can be retrieved from Table 1 of \citet{francois2004}.
\citet{francois2004} showed that concerning the yields from Type Ia SNe, revisions in the theoretical yields by \citet{iwamoto1999} are suggested for Mg:  with the aim of preserving the observed pattern of [Mg/Fe] vs. [Fe/H] they also needed to increase the Mg yields from type Ia SNe by a factor of 5. 
The prescription
for single low-intermediate mass stars is from \citet{van1997}, for the case of the mass loss parameter 
which varies with metallicity \citep[see][model5]{chiappini2003}.
  The choice of such  ad-hoc  nucleosynthesis  prescriptions is supported by the fact that   stellar yields are  still a relatively uncertain component of chemical evolution models \citep{romano2010}.  

  The set of yields used in this paper has been adopted in several works \citep{cescutti2007, spitoni2009, spitoni2011, Mott2013, spitoni2015, spitoni2017}, and turned out to be able to reproduce the main features of the solar neighbourhood.
\section{Model results for the revised ``two-infall'' model}\label{s:revised}
First, we show the results related to the revised ``two-infall'' model
considering the new constraints provided by the stellar ages computed with asteroseismology in terms of the [$\alpha$/Fe] versus [Fe/H] abundance ratios. We recall that with $\alpha$, here we mean the sum of the abundances of Mg and Si. We have adopted the photospheric values of \citet{asplund2005} as our solar reference to be consistent with the APOGEE data release \citep{gperez2016}.

\begin{figure}
\begin{centering}
\includegraphics[scale=0.48]{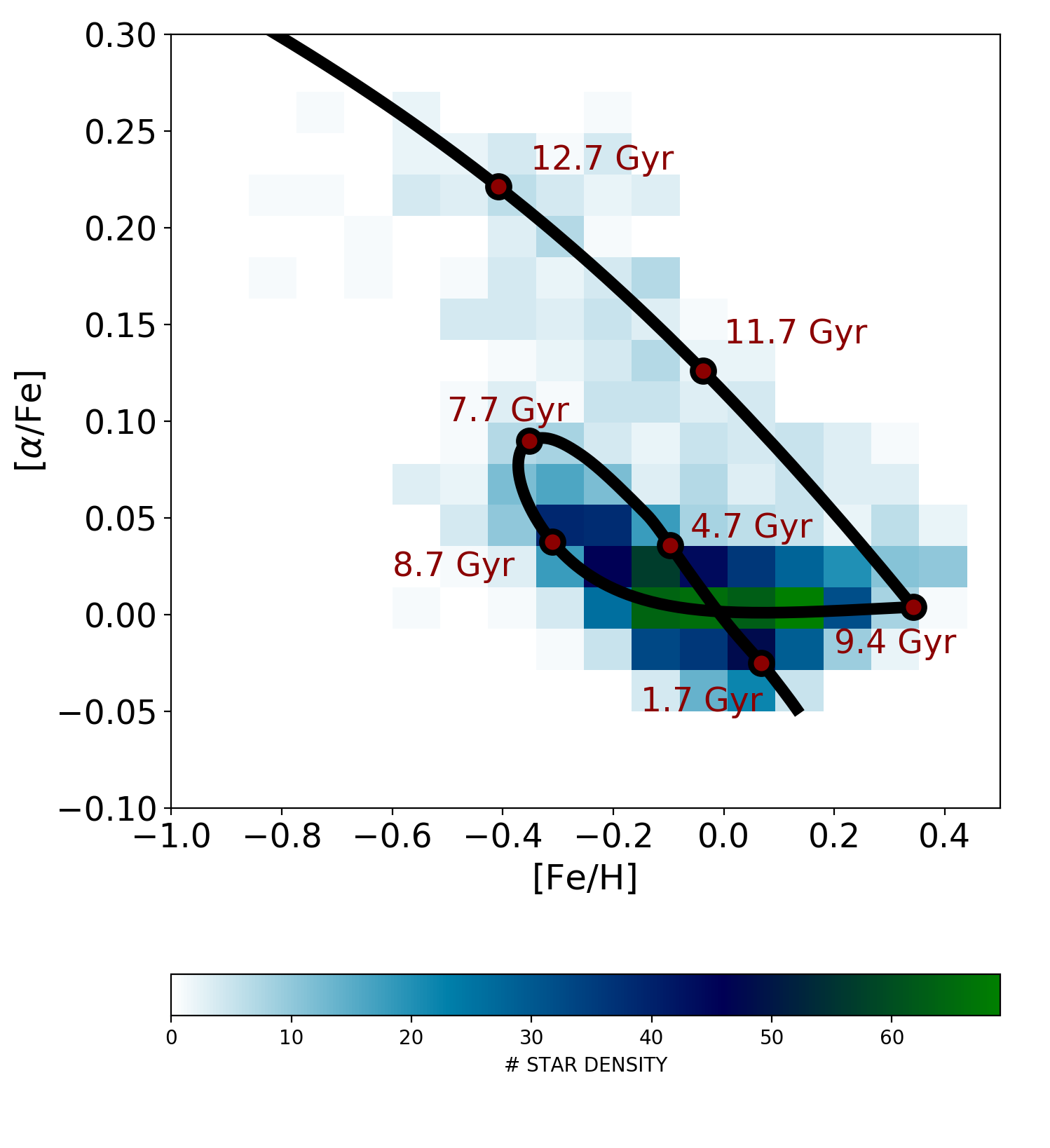}
\caption{The observed density of stars in the [$\alpha$/Fe] vs. [Fe/H] space for the APOKASC stars
by VSA18, compared with our chemical
evolution model (black solid line) in the solar neighbourhood. 
Filled red circles indicate the abundance ratios of the chemical
evolution model at the given age. The area of each bin is fixed at the value of (0.083~dex)$\times$(0.02~dex).}
\label{AFEFE2}
\end{centering}
\end{figure}

We tested several values for the input parameters of the chemical evolution model, and retained the combination that provided the best fit on chemistry and age to the observations reported by VSA18. The parameters included in this model are the following: the first infall is characterized by a current total surface mass density of 8 M$_{\odot}$ pc$^{-2}$  ($\Sigma_{tot1}(t_G)$ in eq. \ref{c1}) and an infall time-scale of $\tau_{D1}$=0.1 Gyr. The second infall corresponds to a current total surface mass density of 64 M$_{\odot}$ pc$^{-2}$  ($\Sigma_{tot2}(t_G)$ in eq. \ref{c2}) with an infall time-scale of $\tau_{D2}$=8 Gyr, and occurs after a delay of $t_{max}=4.3$~Gyr. The star formation efficiency (SFE) is fixed at the value of  $\nu$ =1.3 Gyr$^{-1}$.
The total surface density for the low-$\alpha$ sequence is in agreement with the range of 54 and 74 M$_{\odot}$ pc$^{-2}$ for the thin disc given by \citet[][and references therein]{nesti2013}. For the thick disc surface mass density, the value suggested by \citet{nesti2013} is $\sim$ one tenth of that for the thin disc. In our model we assumed values consistent with this ratio (ratio of total mass surface densities of high-$\alpha$ and low-$\alpha$ is 0.125).

In Fig. \ref{AFEFE1} we compare our best chemical evolution model with
the VSA18 data in the [$\alpha$/Fe] vs. [Fe/H] plane, where an excellent agreement between the two is clearly seen. This level of agreement is only achieved if the second episode of gas infall (related to the low-$\alpha$ sequence) begins when the model curve of the first infall already covers the region populated by some of the stars from the low-$\alpha$ sequence. For this reason, we assume a slightly larger total surface mass density for the high-$\alpha$ component (8 M$_{\odot}$ pc$^{-2}$ than the 6.5 M$_{\odot}$ pc$^{-2}$ adopted by \citet{grisoni2017}), and require a delay time of $t_{max}=4.3$~Gyr for the start of the second episode of gas infall. The sum of the high-$\alpha$ and low-$\alpha$ surface mass densities in our model is, however, very similar to the one of thick and thin disc components of \citet{grisoni2017}.
In Fig. \ref{AFEFE1} we also show a model with the same parameters as in our fiducial one but with a delay time of $t_{max}=1.3$ Gyr, hence similar to the one adopted in the "classical" two infall models  \citep[i.e.][]{chiappini1997,spitoni2009,grisoni2017}. 
It is clear that the high-$\alpha$ stars are not reproduced.
Here we can conclude that the usual delay  adopted  by  the  classical  two infall  model  does  not properly apply to the new VSA18 stellar sample.
In Fig. \ref{AFEFE2} we present the $\alpha$ versus [Fe/H] plane color coded by density of stars, and labeled the ages of the stars created during the
chemical evolution model curve. The densest regions are, as expected, also the regions where our model spends most of the time during its evolution (for roughly 10 Gyr the Galactic disc model is confined to the low-$\alpha$ evolutionary sequence).

Different time-scales of accretion are motivated by the fact that at early times the Galaxy assembled very fast and efficiently, while at later times the formation of the Milky Way proceeded on much longer time-scales as a consequence of dissipative collapse effects \citep{larson1976, cole2000}.

 In the title of this Section we defined our ``two-infall'' model as a ``revised'' one. The novelty of our model compared to the classical ``two-infall'' model by \citet{chiappini1997}, \citet{spitoni2016, spitoni2018} is the long delay before the starting of the second infall of gas.

In Fig. \ref{AFEFE2} the dilution effect caused by the second infall of primordial gas can be appreciated. Contrary to previous models \citep[e.g., those of][]{francois2004, cescutti2007,romano2010,grisoni2017}, the delay of $t_{max}$=4.3 Gyr in the peak of the second infall produces the nearly horizontal stripe at nearly constant [$\alpha$/Fe] from [Fe/H]$\simeq$0.35 dex to [Fe/H]$\simeq$-0.2 dex. The late accretion of pristine gas has the effect of decreasing the metallicity of each stellar population born immediately after the infall event, and has little effect on the [$\alpha$/Fe] ratio since both $\alpha$ and Fe are diluted by the same amount.

When star formation resumes, Type II SNe produce a steep rise in the [$\alpha$/Fe] ratio which is then decreased and shifted towards higher metallicities due to pollution from Type Ia SNe. This sequence produces a loop in the [$\alpha$/Fe] versus [Fe/H] plane of the chemical evolution track, which nicely overlaps with the region spanned by the observed low-$\alpha$ population.
In our picture  the observed ``high-$\alpha$'' sequence can be explained in terms  of our first infall phase, whereas the ``low-$\alpha$'' is reproduced by the second gas infall.

\begin{figure}
\begin{centering}
\includegraphics[scale=0.58]{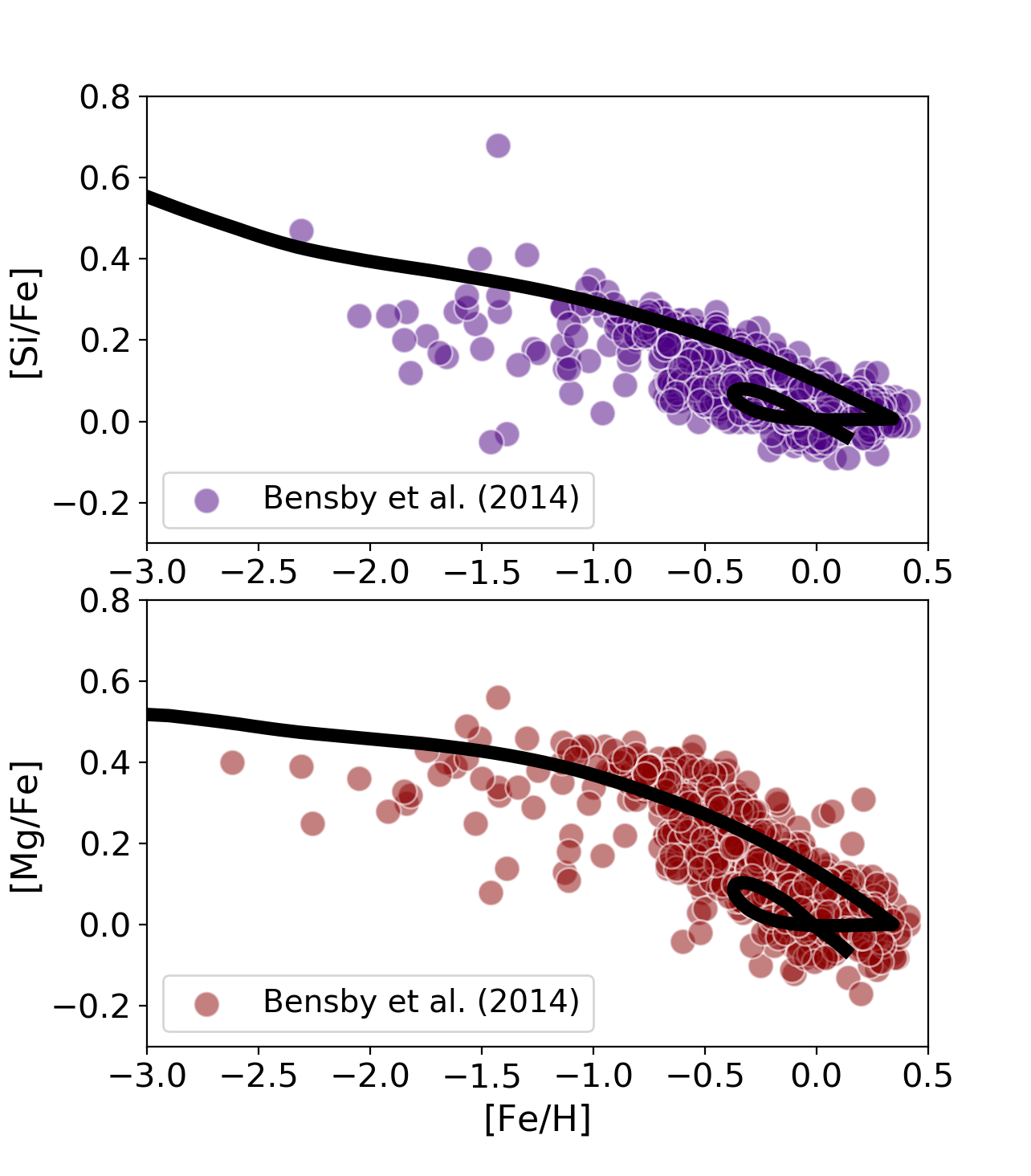}
\caption{Comparison of our updated ``two-infall'' model with the
  observational data for thin and thick disc stars presented by \citet{bensby2014}. {\it Upper panel}: [Si/Fe] vs. [Fe/H]. {\it Lower panel} [Mg/Fe] vs. [Fe/H]. See text for details.}
\label{bensby}
\end{centering}
\end{figure}
\begin{figure}
\begin{centering}
\includegraphics[width=\columnwidth]{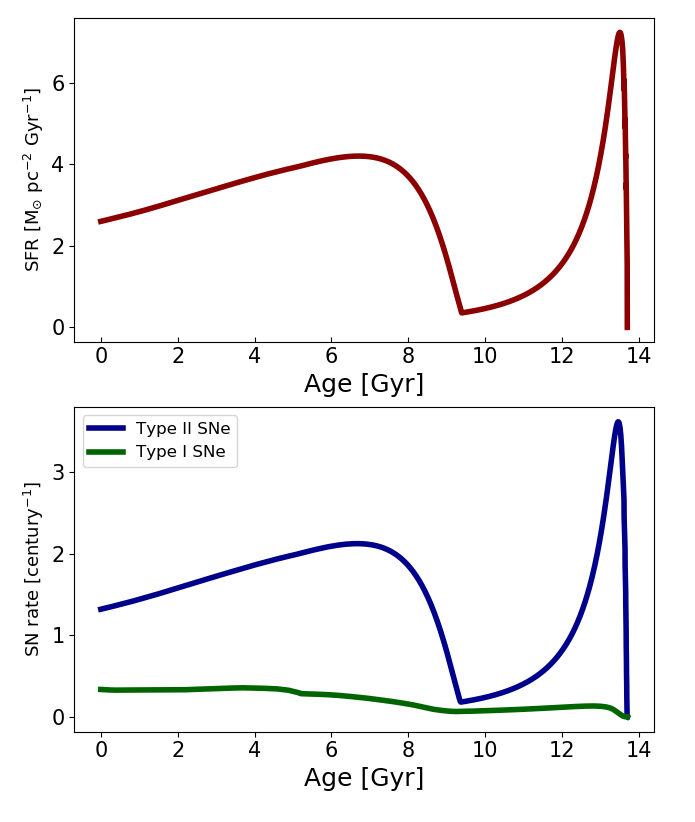}
  \caption{{\it Upper panel}: the SFR time evolution predicted by the
 ``two-infall'' model for the solar neighbourhood. 
 {\it Lower  panel}: the time evolution of the Type Ia SN (blue) and Type II SN (red) rates predicted by the ``two-infall'' model for the whole Galactic
    disc.}
\label{SFR_SN}
\end{centering}
\end{figure}

\begin{figure}
\begin{centering}
\includegraphics[scale=0.45]{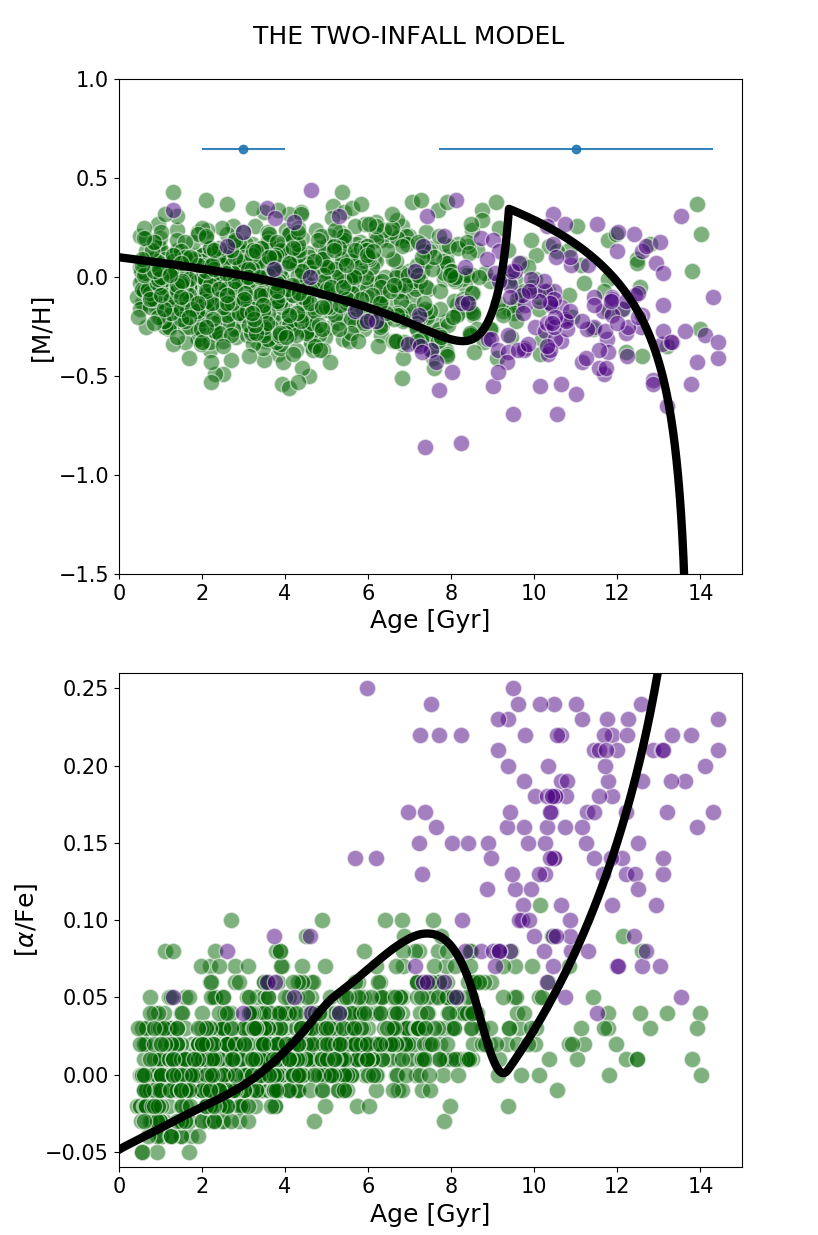}
 \caption{Time evolution of 
 [M/H] (upper panel) and [$\alpha$/Fe]  (lower panel) ratios for the stellar sample presented by VSA18,  compared with our chemical evolution model predictions (black
 solid line).  Purple filled circles depict the high-$\alpha$ population,
 whereas green filled circles represent the low-$\alpha$ one.
In the upper panel light blue symbols indicate median age uncertainties for 3 and 11 Gyr old stars of the VSA18 sample.
}\label{AFE_age_SD}
\end{centering}
\end{figure}
\begin{table}[htp]
\begin{center}
\caption{Theoretical and observed solar abundances. }
\label{tab_02}
\begin{tabular}{c|cc}
  \hline
%\noalign{\smallskip}
\\
 Element abundance &{\it Observations} &  {\it Model}\\
 $\log$($X$/H)+12 & Asplund et  al. (2005)&\\
 & [dex]&[dex]\\
%\noalign{\smallskip}
%\noalign{\smallskip}
\hline
%\noalign{\smallskip}
Fe &7.45$\pm$0.05&7.40 \\
\\
%\noalign{\smallskip}
 \hline
%\noalign{\smallskip}
Si &7.51$\pm$0.04& 7.47  \\
\\
%\noalign{\smallskip}
 \hline
%\noalign{\smallskip}
Mg& 7.53$\pm$0.09 &7.49  \\
\\
%\noalign{\smallskip}
 \hline
%\noalign{\smallskip}
\end{tabular}
\end{center}
\end{table}

\begin{figure}
\begin{centering}
\includegraphics[scale=0.44]{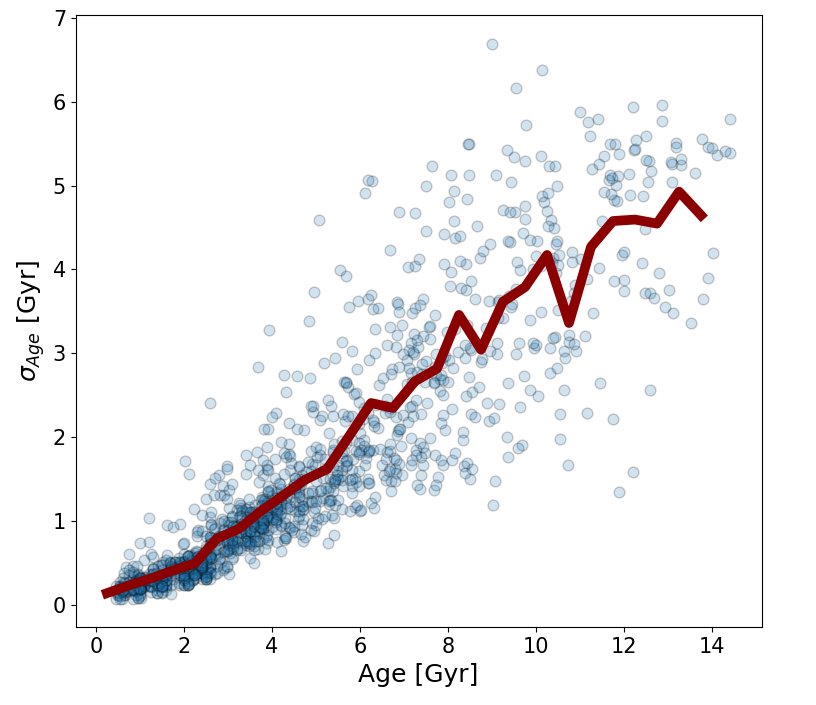}
  \caption{Observed error estimate on the stellar ages for the whole data sample (both low-$\alpha$ and high-$\alpha$ stars) by VSA18 (blue circles). The red line is the average observed error estimate on age bins of 0.5 Gyr.}
\label{errors}
\end{centering}
\end{figure}

\begin{figure*}
\begin{centering}
\includegraphics[scale=0.55]{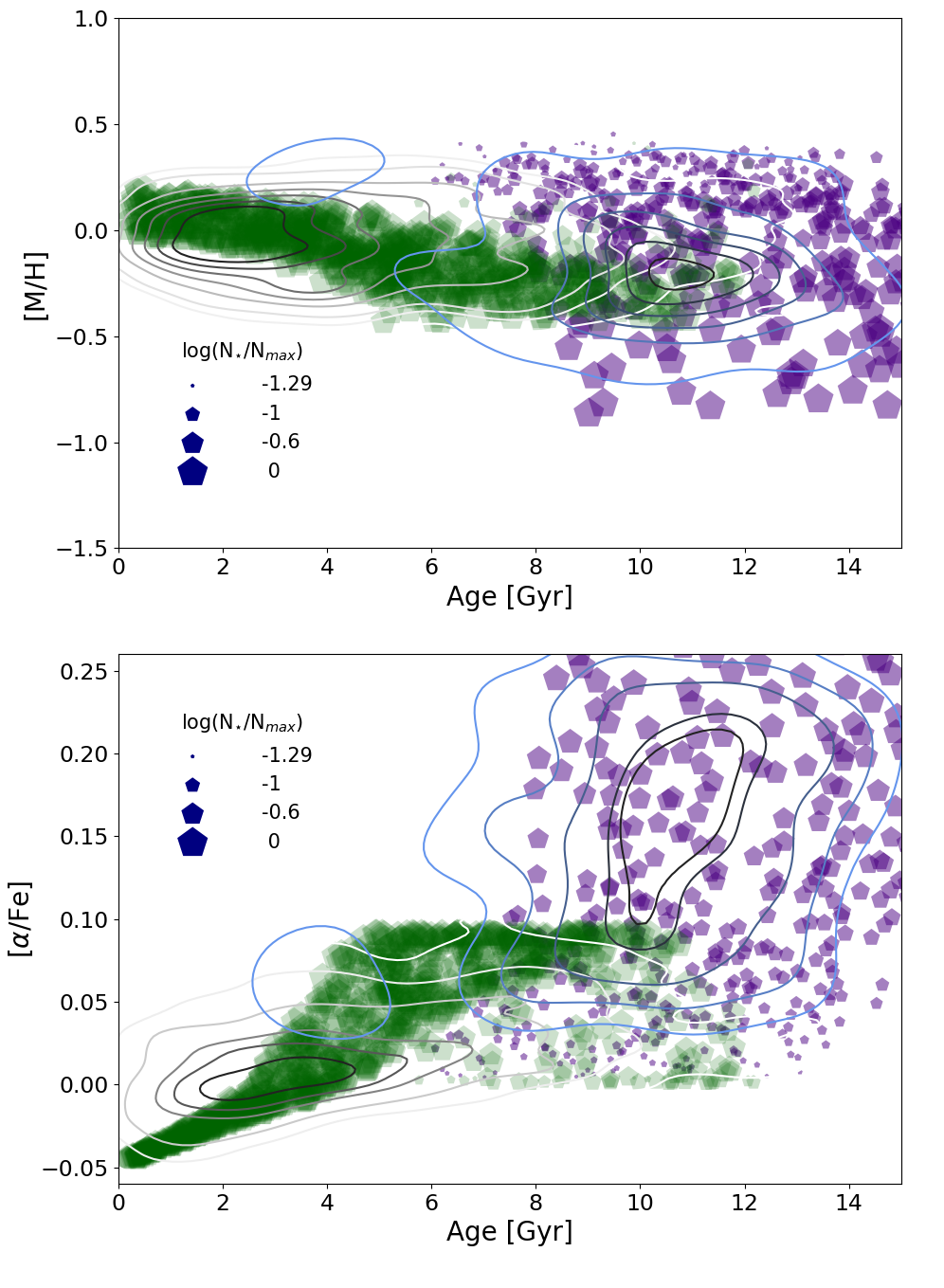}
  \caption{ Chemical evolution model results for the time evolution of [M/H]  (upper panel) and [$\alpha$/Fe] (lower panel)  distributions produced from the age and metallicity uncertainties reported by VSA18 following Eqs.~\ref{Eqer1} and~\ref{Eqer2}. 
 The purple filled pentagons are the  stars formed during the first infall ( ``high-$\alpha$'' sequence),  whereas the green filled pentagons are the  ones   formed in ``low-$\alpha$'' sequence. 
The pentagon size indicates the the  local number density of formed stars normalized to maximum value. We also show the contour density curves of the observed ``high-$\alpha$'' stars (black and blue lines stand for the higher and lower levels, respectively) and ``low-$\alpha$'' stars (black and white lines stand for the higher and lower levels, respectively) of the VSA18 sample.   }
\label{sdens}
\end{centering}
\end{figure*}

A long delay between the gas infall episodes has been reported in simulations of late-type galaxies within a cosmological framework. For instance, \citet{calura2009} investigated the chemical properties of Milky Way-like galaxies using a semi-analytical model within the hierarchical picture of galaxy formation and predicted the presence of an horizontal stripe in the [O/Fe] versus [Fe/H] plane  caused by the presence of a substantial increment of late infall episodes.
 Moreover, \citet{grand2018} studied the stellar disc properties of different Milky Way sized haloes extracted from very high resolution cosmological zoom-in AURIGA simulations. They found that a bimodal distribution in the [$\alpha$/Fe] vs. [Fe/H] plane is present when an early high-[$\alpha$/Fe] star formation phase in the disc is followed by a shrinking of the gas disc. This shrinking is caused by a temporarily significant lowered gas accretion rate at ages between 6-9 Gyr, after which disc growth resumes through the occurrence of another infall episode. In our ``two infall'' model  a lowering of the gas accretion is mimicked by a consistent delay in the second infall of gas.

A late time second accretion phase in a two-infall context has been also derived by \citet{noguchi2018}  using the   ``cold flow''  model by  \citet{dekel2006} for cold-flows/shock heating. 
The Milky Way has been simulated   using a code that divides the Galactic disc  into a series of concentric annuli and the growth of the  virial mass of the dark matter halo  follows  cosmological numerical simulations  by \citet{Wechsler2002}.

A first infall episode originates the high-$\alpha$ sequence, which is followed by a hiatus of 2 Gyr until the
 shock-heated gas in the galactic dark matter halo has radiatively cooled and can accrete onto the Galaxy. The low-$\alpha$ sequence stars form during a second phase, after the hiatus. In agreement with our model, the SFR is characterized by two peaks, and in \citet{noguchi2018}  they  are separated by around five billion year.
However, the pause in gas infall in \citet{noguchi2018}  is present at later times  than the one presented here, due to the fact that the author did not use the constraint  from stellar ages to locate the infall hiatus as done in our analysis.

In Table~\ref{tab_02}, the solar abundances of Fe, Mg and Si predicted by our ``two-infall'' model are compared with observations. The model solar abundances are determined from the composition of the ISM at the time of the formation of the Sun (after 9.5 Gyr from the Big Bang). It is evident that our model is able to reproduce the solar abundance ratios for the elements considered in this work.
We have also tested if our model is capable of reproducing the trends in Mg and Si provided by other observational studies and surveys. To this aim in Fig \ref{bensby} we compare the  [Mg/Fe] and [Si/Fe] vs [Fe/H] predicted by our model in the solar neighbourhood, with the Gaia-ESO data reported by \citet{bensby2014}. We find that our model  reproduces satisfactorily the Gaia-ESO data.

An important constraint s is represented by the present-time SFR in the solar vicinity. The  upper panel of Fig. \ref{SFR_SN} shows the time evolution of the SFR in our model, which predicts a present day  SFR value of 2.60 M$_{\odot}$ pc$^{-2}$ Gyr$^{-1}$. This is in excellent agreement with the measured range in the solar vicinity of 2-5 M$_{\odot}$ pc$^{-2}$ Gyr$^{-1}$ as suggested by \citet{matteucci2012} and \citet{prantzos2018}. 

By adopting a closed-box chemical evolution model, in \citet{Snaith:2015ki} and \citet{Haywood:2016el}   the SFR has been selected ad-hoc in order to reproduce the solar neighbourhood chemical  data by  \citet{Adibekyan2012} and \citet{haywood2013}. This different methodology compared to the two-infall approach leads roughly to the same transition time between the two $\alpha$ sequences, supporting  our results and the robustness of the updated two-infall model. 
  
The time evolution of the Type Ia SN and Type II SN rates is plotted in the lower  panel of Fig.~\ref{SFR_SN}. The present day Type II SN rate in the whole Galactic disc predicted by our model is 1.31 /[100 yr], in good agreement with the observations by \citet{li2010} which yield a value of 1.54 $\pm$0.32 /[100 yr]. 
The predicted present day Type Ia SN rate in the whole Galactic disc
is 0.33 /[100 yr], slightly  below the observations by \citet{li2010} which yield 0.54$\pm$0.12 /[100 yr], but in excellent agreement with the value provided by \citet{cappellaro1997} of 0.30$\pm$0.20  /[100 yr].

\begin{figure*}
\begin{centering}
\includegraphics[scale=0.43]{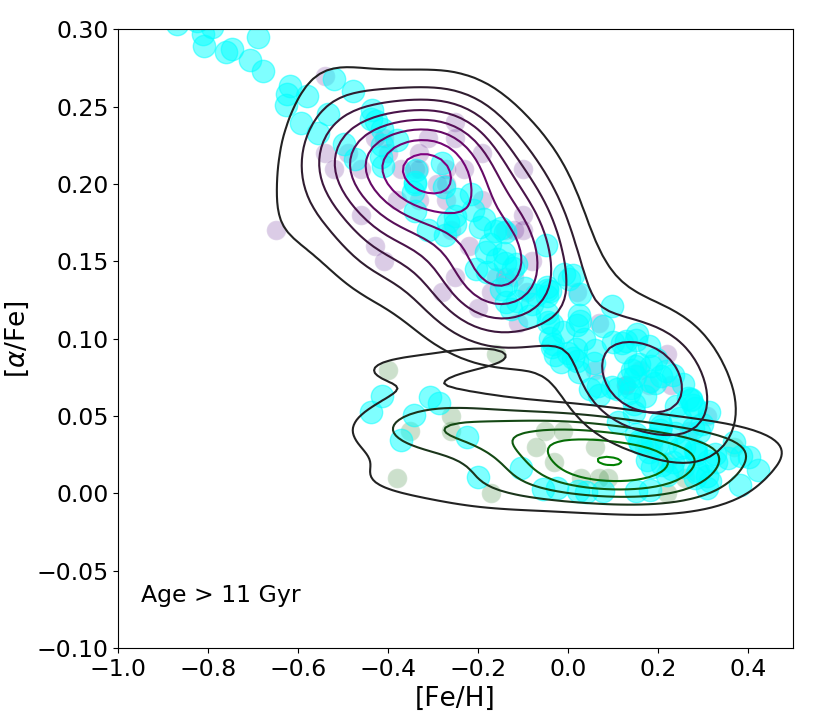}
\includegraphics[scale=0.43]{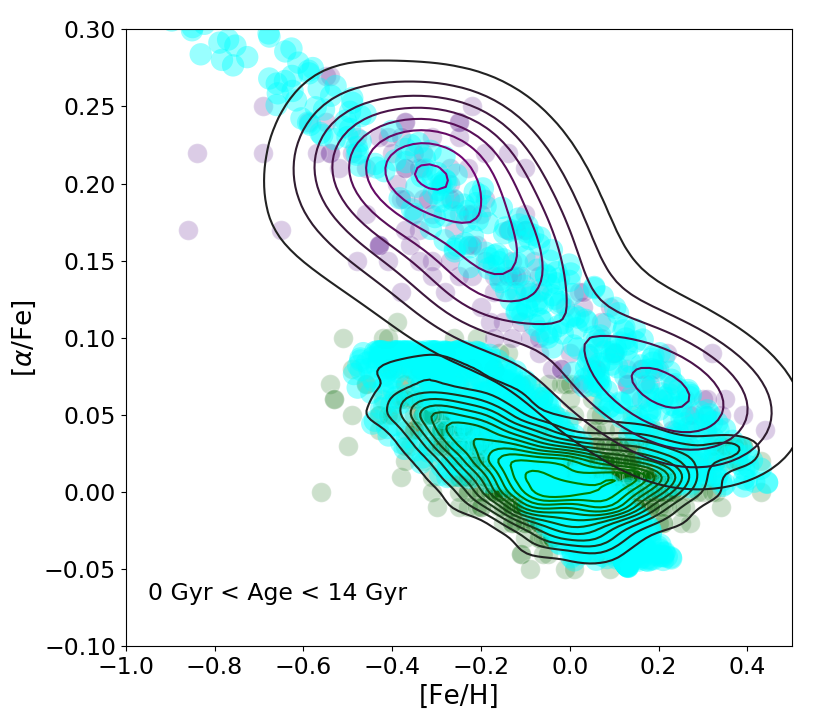}
\includegraphics[scale=0.43]{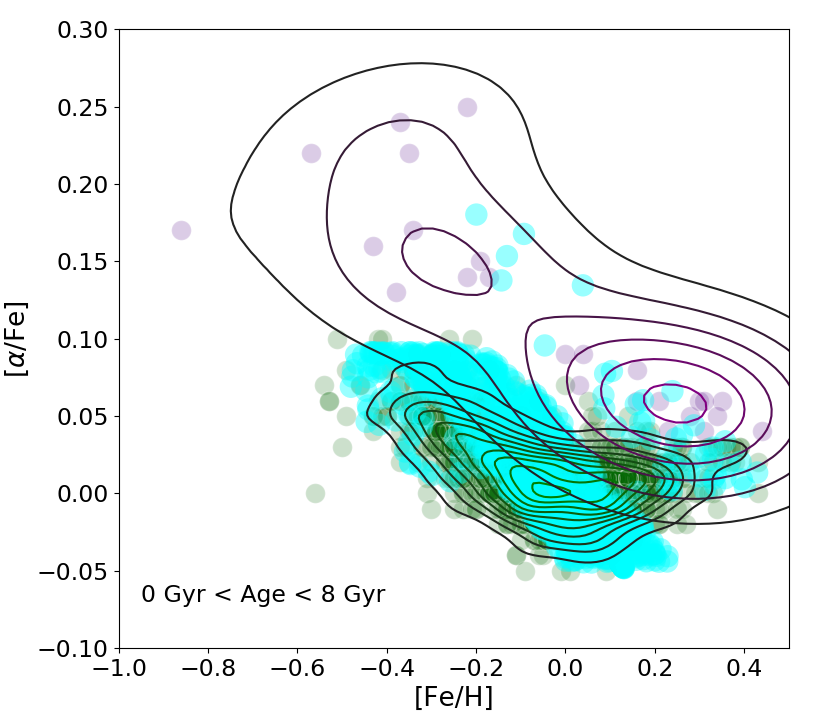}
\includegraphics[scale=0.43]{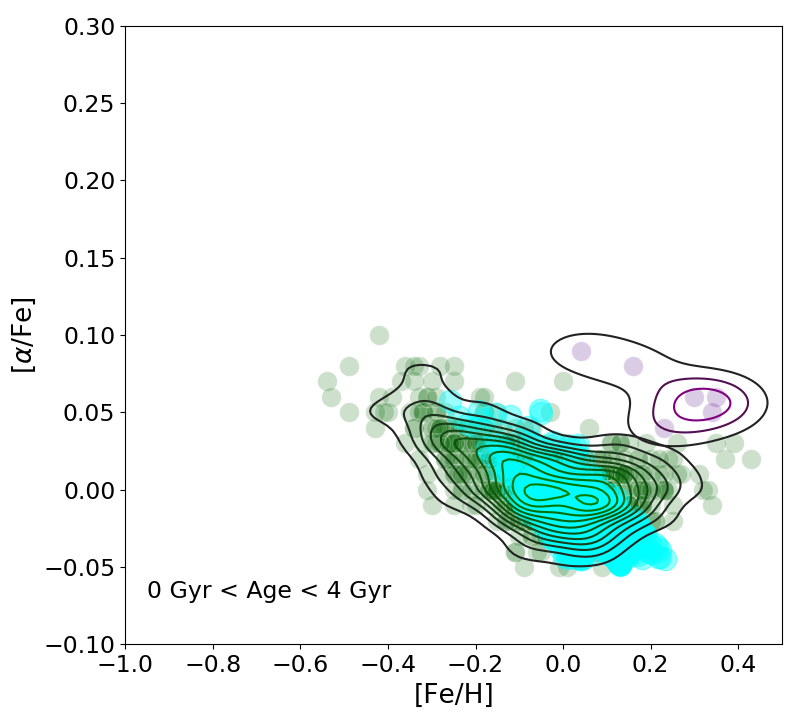}
\caption{The [$\alpha$/Fe]  abundance ratios as a
  function of [Fe/H] predicted by our ``two-infall'' chemical evolution model (cyan circles) taking into account the average observational errors on age and metallicity (following Eqs.~\ref{Eqer1} and~\ref{Eqer2}) computed for different age ranges. Also plotted are the observational data, color coded  as in Fig.~\ref{AFEFE1}.
 {\it Upper Left Panel}:  stars older than 11 Gyr. 
{\it Upper Right Panel}: stars formed throughout the Galactic history.
{\it Lower Left Panel}: stars younger than 8 Gyr.
{\it Lower Right Panel}: stars younger than 4 Gyr.
}
\label{AGES_AFE_FEH}
\end{centering}
\end{figure*}

\begin{figure}
\begin{centering}
\includegraphics[scale=0.38]{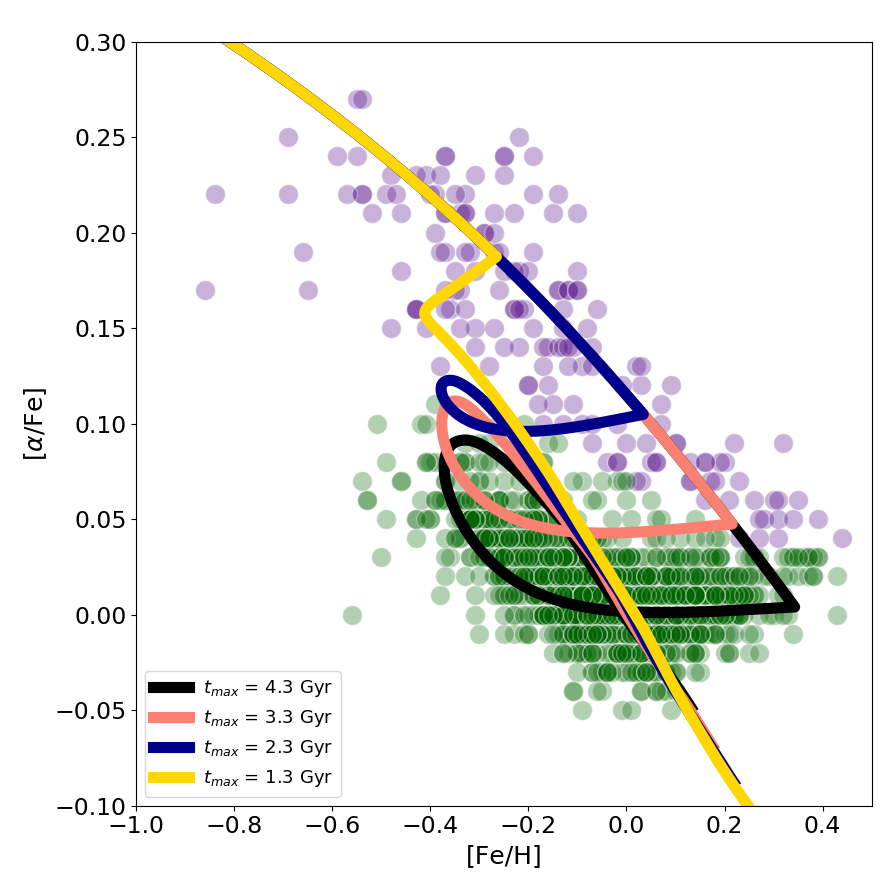}
\includegraphics[scale=0.38]{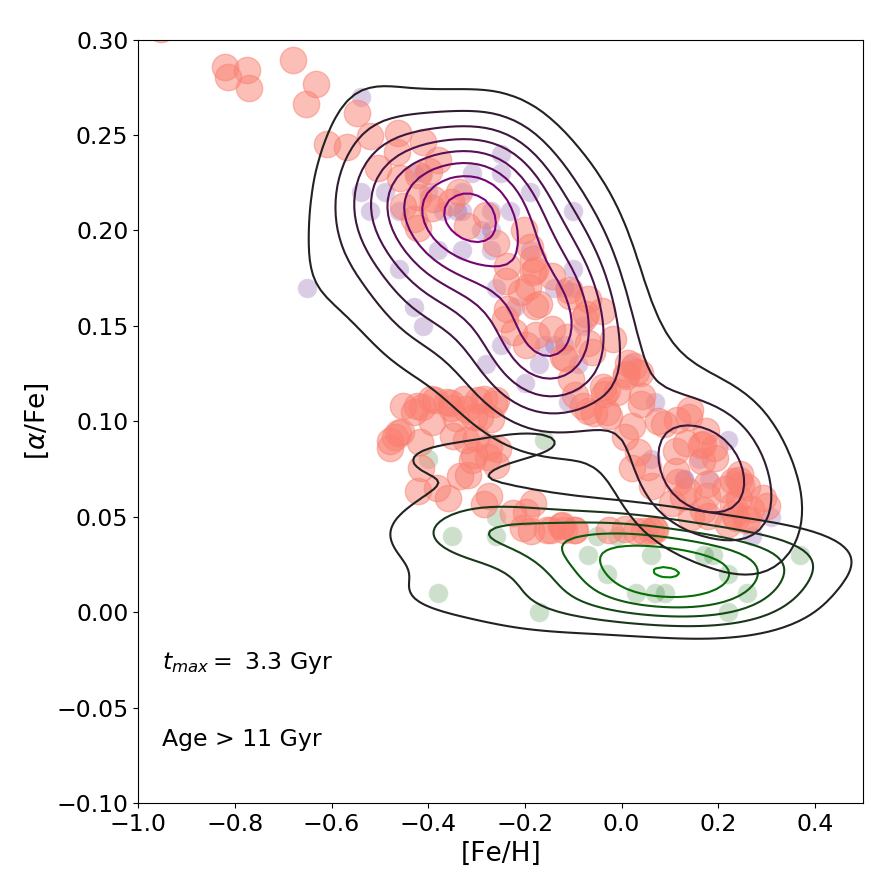}
\caption{{\it Upper  Panel}:  Effects on the chemical evolution of the solar neighbourhood in
  the [$\alpha$/Fe] versus [Fe/H] abundance ratios  of different values for   the delay of the beginning of the second infall (the quantity $t_ {max}$ in eq. \ref{a}).
 {\it Lower  Panel}: The [$\alpha$/Fe]  abundance ratios as a
  function of [Fe/H] predicted by our ``two-infall'' chemical evolution model (salmon circles) taking into account the average observational errors on age and metallicity (following Eqs.~\ref{Eqer1} and~\ref{Eqer2})  for stars older than 11 Gyr, considering a delay for the beginning of the second infall of $t_ {max}$= 3.3 Gyr. Also plotted is the observational data, color coded is as in Fig.~\ref{AFEFE1}.
}
\label{tmax}
\end{centering}
\end{figure}

In the VSA18 sample, the metallicity [M/H] is computed using the following  expression introduced by \citet{salaris1993}:
\begin{equation} 
 \mbox{[M/H]}   = \mbox{[Fe/H]} (t)  + \log \left( 0.638 \times
   10^{[\alpha/Fe]}+ 0.362 \right).
\label{MH}
\end{equation}
We combined the abundance ratios [Fe/H] and [$\alpha$/Fe] predicted by our model using this formulation to be consistent with the data. In Fig. \ref{AFE_age_SD} the results of the time evolution of the metallicity [M/H] and the [$\alpha$/Fe] ratios are reported. Also here the effect of the dilution is evident, which produces a drop in the [M/H] when the second infall takes place. The general trend of the data is reproduced, however the VSA18 sample does not seem to show this kind of "knee" feature at an age of $\sim$9.6 Gyr.

In the lower panel of  Fig. \ref{AFE_age_SD} the time evolution of
[$\alpha$/Fe] is presented. When the second infall begins, a sudden increase of [$\alpha$/Fe] is predicted by our model as a result of the accretion of new primordial gas: Type II SNe (which trace the SFR) can pollute the ISM with
$\alpha$ elements on short times scales while Type Ia SNe need longer time-scales to substantially pollute the ISM with Fe. The expected decrease in the [$\alpha$/Fe] abundance is seen $\sim$2~Gyr after the second infall.

In conclusion, our revised two-infall model can reproduce the main features presented in the VSA18 dataset. The sudden drop in [M/H] and increase in [$\alpha$/Fe] associated to the second accretion episode are not obvious in the observations but can be hidden behind the observational uncertainties. In  next Section we will present the ``two-infall''  model results taking into account the  error estimates related to stellar ages and metallicity. 
\section{Model results taking into account  the observational errors}
In Fig. \ref{errors} we report the average errors in bins of 0.5 Gyr as a function of the Galactic age for the estimated  stellar ages for the APOKASC sample by VSA18. 

We  note that the errors in the stellar age determination are strongly
dependent on the Galactic age, and  span  a huge range of values: between $\sigma_{Age}$=0.13 Gyr (at the Galactic age of 0.25 Gyr) and $\sigma_{Age}$=4.93 Gyr (at a Galactic age of 14 Gyr). On the other hand, the errors on the abundance ratio [M/H] reported by APOGEE are independent from the stellar ages and the average value is $\sigma_{[M/H]}$ $\sim$ 0.118 dex.

We take into account these errors in our model by adding, at each Galactic time, a random error to the ages and metallicities [M/H] of the stellar populations
formed at Galactic time $t$. These random errors are uniformly distributed in the interval described by the average errors estimated at that time (see solid red line in Fig.~\ref{errors}), and we define the ``new age'' including these uncertainties as follows:
\begin{equation} 
 \mbox{Age}_{new}(t)   = \mbox{Age} (t)  +  U([-\sigma_{Age}(t), \sigma_{Age}(t) ]),
\label{Eqer1}
\end{equation}
where  Age $(t)$ = (13.7- $t$) Gyr, and $U$ is the random generator function.
Similarly, we implement the error in the chemical abundance space through the following relation:
\begin{equation} 
 [\mbox{M/H}]_{new}(t) =  [\mbox{M/H}](t)+  U([-\sigma_{[M/H]}(t), \sigma_{[M/H]}(t)
 ]).
\label{Eqer2}
\end{equation}

In  Fig. \ref{sdens} the results of the
metallicity time evolution [M/H] and the [$\alpha$/Fe] ratios including the
errors described in eqs. (\ref{Eqer1}) and  (\ref{Eqer2}) are reported. With the purple filled pentagons we label the stellar populations formed at Galactic time $t<t_{max}$ =4.3 Gyr (time for the maximum infall on the  second gas infall) before including the errors. Analogously, the  filled pentagons  green circles represent the
stellar populations formed at Galactic time $t>t_{max}$ =4.3 Gyr.

In the remainder of the paper we will refer to our chemical evolution model
combined with the observational errors as to our {\it synthetic model}.
Comparing our model results with the data, we see that they nicely
reproduces the observational trends, and our synthetic model results do not show the "knee"-like feature in the [M/H] and  [$\alpha$/Fe] associated to the second gas infall episode (in contrast to the model curves in the two upper panels of the same figure). The inclusion of the error in the stellar ages and metallicites allows our model to also predict a presence component of old (t$>$10~Gyr) low-$\alpha$ stars that are observed in the dataset. Moreover it can be seen that, in the [$\alpha$/Fe]  vs. age plot, the observed spread of the oldest high-$\alpha$ stars is comparable with the one of the model.

 Our ``simple'' scenario is able to explain the general trends of the APOKASC sample.
We underline that the sample presented by VSA18 selected stars with a [M/H] $>$-1 dex, and therefore we show  only predicted  stars   above this threshold for  [M/H].
  With  variable symbol size we also indicate the density of  the local number of  formed stars  predicted by our synthetic model normalised to its maximum number.
In  upper panel  of Fig. \ref{sdens}  it is visible  that some of the high-$\alpha$ stars predicted by our  synthetic  model show larger [M/H] values  than the observed ones (purple pentagons outside the blue contour lines). These stars appear in lower panel of Fig. \ref{sdens} at ages older than 6 Gyr and [$\alpha$/Fe] abundance ratios smaller than 0.05 dex, not seen in the VSA18 observations. However, the number density of  these high-$\alpha$ stars is negligible compared to the density of low-$\alpha$ stars formed at the same age. In conclusion, there is no tension between our synthetic model results and the VSA18 sample. We notice that our synthetic high-$\alpha$ stars present less spread than the observed one, and stellar migration from outer Galactic regions could be a possible explanation for this discrepancy \citep[][]{schoenrich2009}.

VSA18 claimed that the majority of the observed high-$\alpha$ stars have ages in the range of $\sim$8 to $\sim$14 Gyr and show no tight correlation between age and [$\alpha$/Fe].  Our synthetic model for the high-$\alpha$ sequence is in agreement with this statement. In fact, the region with higher predicted  stellar density is above [$\alpha$/Fe]=0.05 dex, and the inclusion of observational errors washes away the visible tight correlation in the chemical evolution results.  \citet{RojasArriagada:2016eq} claimed  that the formation  of the high-$\alpha$ and low-$\alpha$ sequences is not strictly sequential but they partly overlap in time. This fact would imply that the two sequences are not entirely sequential as noted in   \citet{haywood2013} and  \citet{noguchi2018}. On the other hand, the inclusion of observational errors in our synthetic model creates in the age-metallicity and in the age-[$\alpha$/Fe] a partial overlapping of the two stellar sequences as visible in Fig. \ref{sdens}.

In Fig \ref{AGES_AFE_FEH} we compare our model results with the inclusion of errors with the observational data in the classical chemical evolution plot [$\alpha$/Fe] vs. [Fe/H] for different stellar ages. Overall, it is clearly shown how our synthetic model including observational errors adopting eqs.~(\ref{Eqer1}) and~(\ref{Eqer2})  fits the data extremely well.
The upper   left panel shows the stars VSA18 sample with ages older than 11 Gyr. It is also evident that the oldest stars seem to confirm our Astroarcheology scenario: they keep the signature of the delayed infall of gas and the successive dilution effect on the [$\alpha$/Fe] vs. [Fe/H] relation as shown by the horizontal stripe at roughly constant [$\alpha$/Fe]. However, stars considered here are older than 11 Gyr, whereas our delayed infall starts 4.3 Gyr after the ``Big Bang'' (corresponding to a Galactic age of 9.4 Gyr). When taking into account the observational errors in our synthetic model, stars born shortly after the second infall episode have corresponding age uncertainties large enough to make them consistent with ages older than 11 Gyr (c.f., Fig~\ref{errors}). Thus, our model suggests that the population of old low-$\alpha$ stars is an artifact created by large errors in the ages of stars older than $\sim$8 Gyr.

In the  upper right panel all stars of the sample are compared with our model, showing that very good agreement between data and model prediction is clearly achieved. The   lower left panel presents ages younger than 8 Gyr, and we note that the bulk of stars predicted by our model belong to the low-$\alpha$ sequence and only a small fraction presents high-$\alpha$ values. The data shows a small number of stars in the high-$\alpha$ sequence at metallicities below [Fe/H]$\simeq$-0.3 not predicted by our model, but the overall agreement is good. Finally, in the  lower right  panel shows the case of stars younger than 4 Gyr. Almost all the observed stars belong to the low-$\alpha$ sequence and occupy a metallicity range larger than that predicted by our model. However, the region between -0.2 dex $<$ [Fe/H] $<$ 0.3 dex where are synthetic results are located corresponds to the area with the highest number of observed stars.
\begin{figure}
\begin{centering}
\includegraphics[scale=0.51]{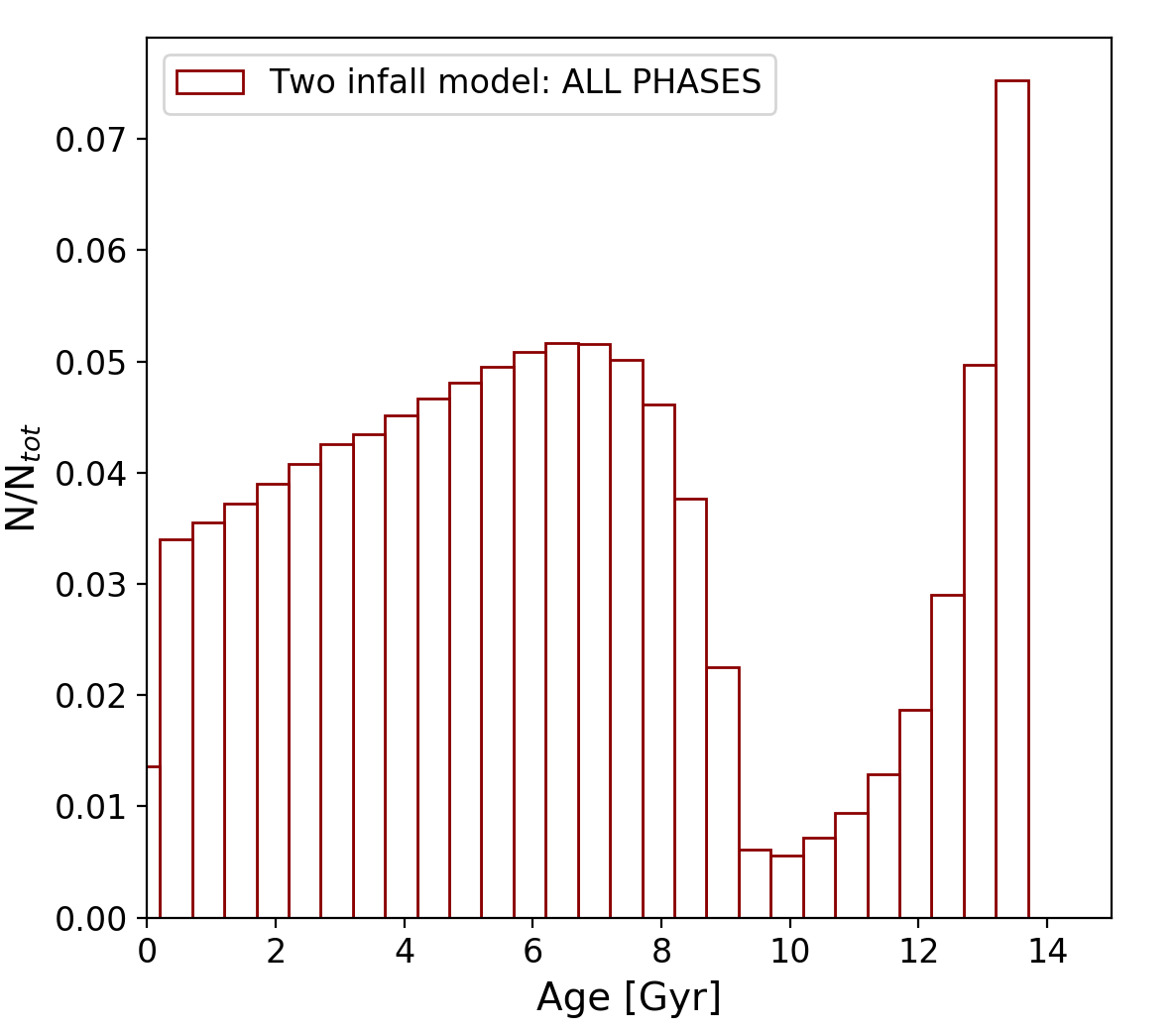}
\includegraphics[scale=0.51]{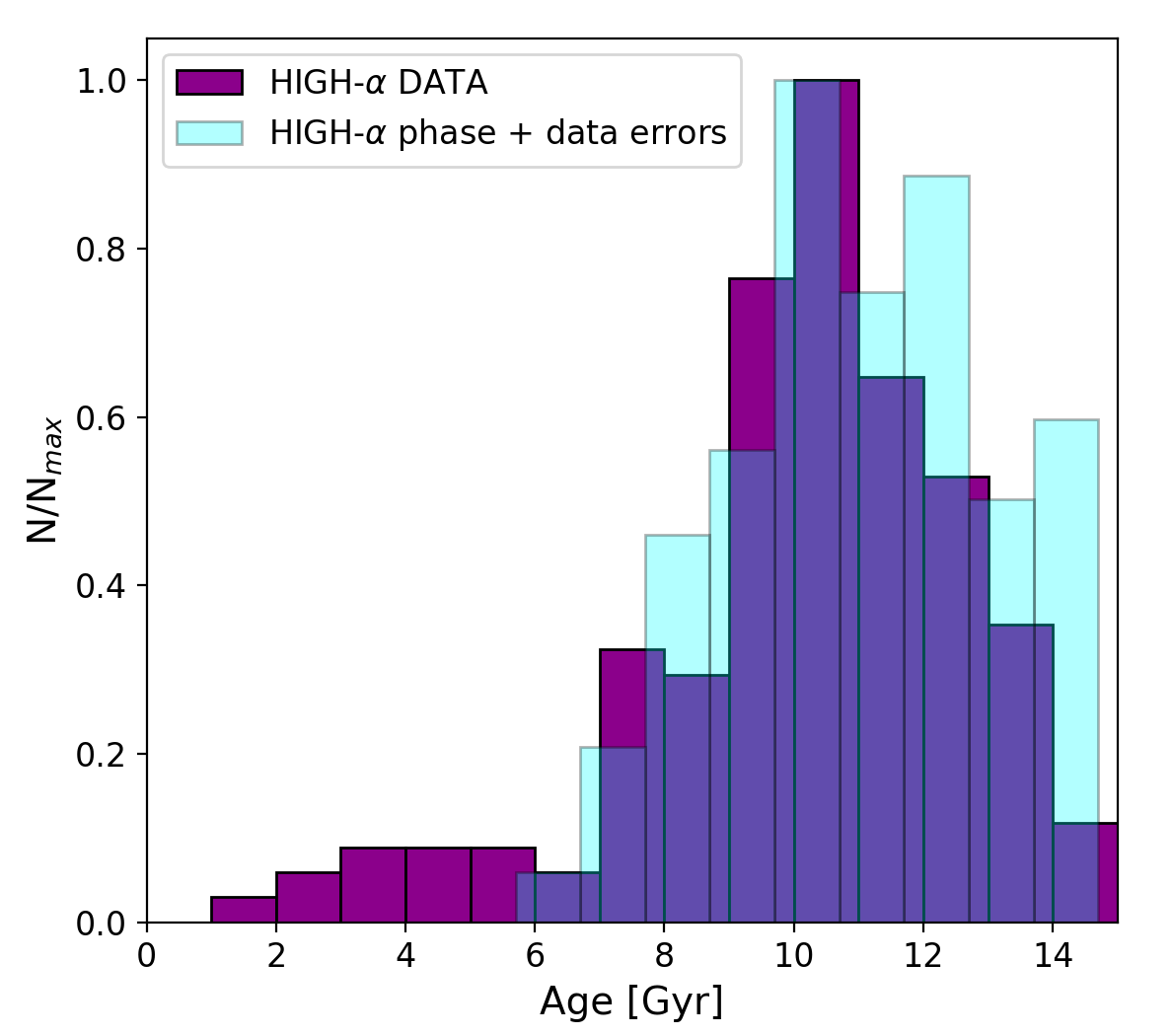}
\includegraphics[scale=0.51]{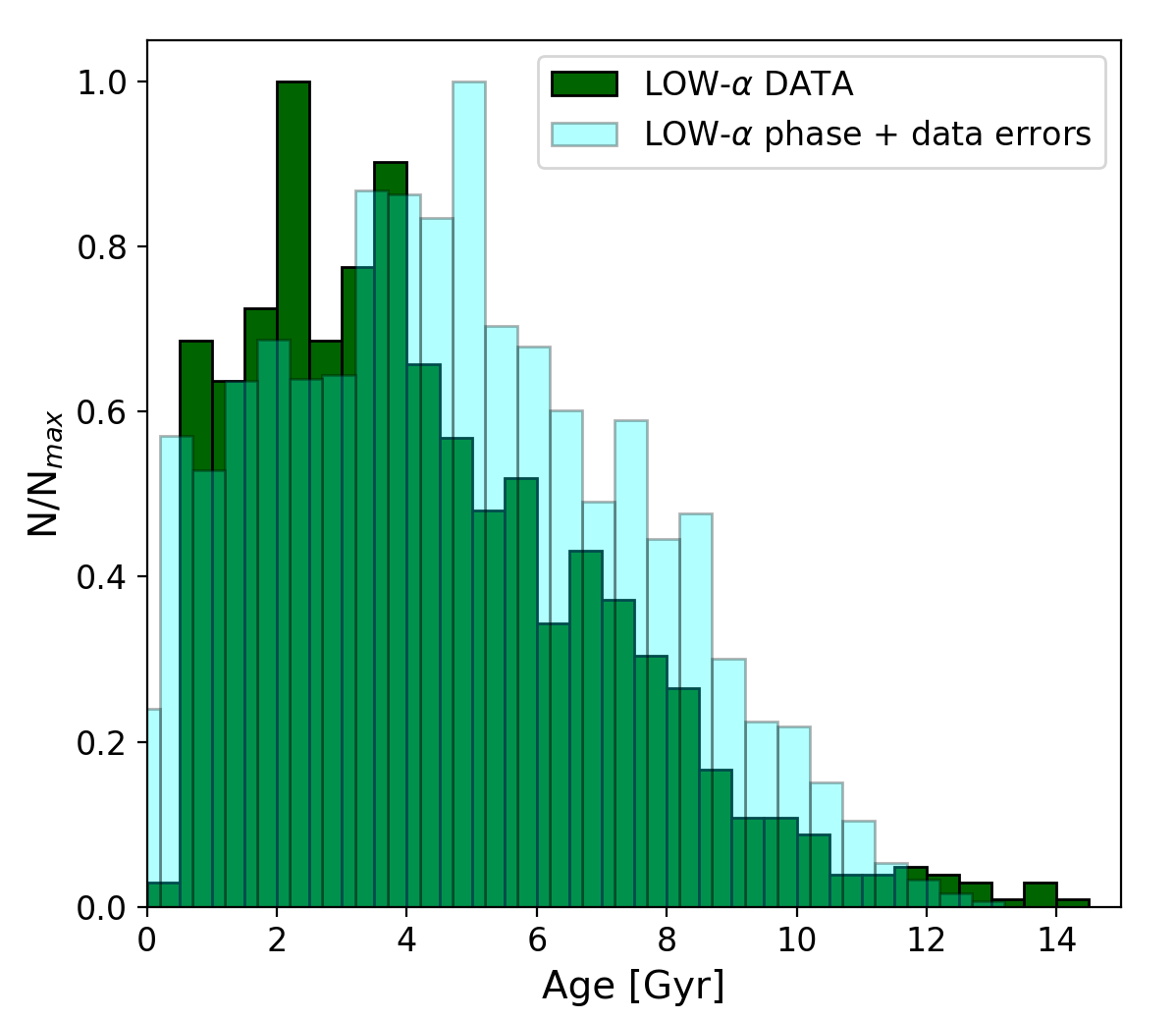}
\caption{
%Stellar age distributions predicted by our ``two-infall''
 %   chemical evolution models.
{\it Upper panel}:
 age distribution of stars formed
    during the whole Galactic
    history predicted by our ``two-infall''
  chemical evolution models.
{\it Middle
      panel}:  age
    distribution of the observed high-$\alpha$ sequence stars  of the
    VSA18 sample (purple histogram) compared to the predicted  distribution of stars
    created during the first infall episode, and taking into account
    the errors given by  eqs. (\ref{Eqer1}) and (\ref{Eqer2})  (cyan histogram).
    {\it Lower panel}: age
    distribution of low-$\alpha$ stars in the VSA18 sample (green histogram)
    compared to the model-predicted distribution of stars
    created during the second infall episode, taking into account
    the errors given by  eqs. (\ref{Eqer1}) and (\ref{Eqer2}) (cyan histogram). 
}
\label{SAGE}
\end{centering}
\end{figure}

 The value of $t_{max}$  has been tuned by imposing that our synthetic model should be able to reproduce the  observed stars older than 11 Gyr (upper left of  Fig. \ref{AGES_AFE_FEH}) in  the $[\alpha$/Fe] vs [Fe/H] relation. 
Including the observational error, our ``best model'' should  predict the horizontal stripe which characterises  low-$\alpha$  sequence stars older than 11 Gyr.

We recall that in this work we adopt for the high-$\alpha$ sequence the same infall time scale  of  the thick disc phase  of   \citet{grisoni2017}, i.e. $\tau_{D1}$=0.1 Gyr. With this particular value  \citet{grisoni2017}  nicely reproduced the metallicity distribution function (MDF) of the thick disk stars  of  the AMBRE project. 
Fig.  \ref{AFEFE1} shows that the high-$\alpha$ sequence stars in the  [$\alpha$/Fe] vs. [Fe/H] relation  of the VSA18 sample are also well reproduced  with  $\tau_{D1}$=0.1 Gyr:    the model line passes through the higher density peaks indicated by the contour plot curves.
In the upper panel of Fig. \ref{tmax},  we explore different $t_{max}$ values spanning  the range between 1.3 and 4.3 Gyr, assuming  $\tau_{D1}$=0.1 Gyr.
It is clear that already    a delay $t_{max}$   1 Gyr shorter than the one adopted in this work will not allow us to reproduce the data. 
In the lower panel of   Fig. \ref{tmax} we show the predicted   [$\alpha$/Fe] vs. [Fe/H] abundance ratios  by the synthetic model for stars older than 11 Gyr with   $\tau_{D1}$=0.1 Gyr, $\tau_{D2}$=8  Gyr and $\tau_{max}$=3.3 Gyr.  The horizontal stripe is located at a much higher  [$\alpha$/Fe] value than  the one shown by  low-$\alpha$ data.

We stress that the location of the horizontal stripe is independent to  the choice of  $\tau_{D2}$. Naturally, different values of $\tau_{D2}$ will lead to different 
 loop sizes in the low-$\alpha$ sequence:  smaller values will lead to more extended loops in the [$\alpha$/Fe] vs [Fe/H] relation (see discussion in Section 6).

\begin{figure}
\begin{centering}
\includegraphics[width=\columnwidth]{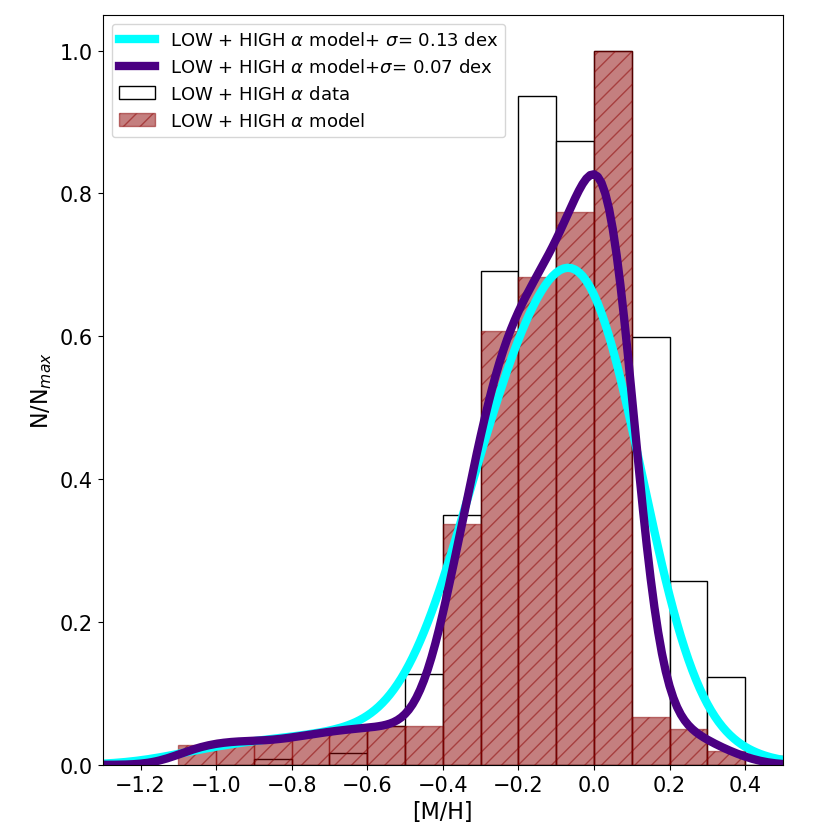}
  \caption{The metallicity distribution predicted by our chemical
    evolution model is indicated by the red histogram. The observed
    distribution  calculated including both  high-$\alpha$ and low-$\alpha$
    stars is shown by the black,
     empty histogram. The cyan
    line indicates the metallicity distribution of our chemical
    evolution model convolved with a gaussian with standard deviation $\sigma=  0.13$~dex. Finally, the purple
    line indicates the metallicity distribution of our chemical
    evolution model convolved with a gaussian with standard deviation $\sigma=0.07$~dex.}
\label{MDF}
\end{centering}
\end{figure}

Finally, the value of $\tau_{D2}$ has been tuned with the aim of: 
\begin{itemize}
 \item reproducing 1) the present day  SFR, 2) the present day Type Ia and Type II SN rates, 3) the solar abundances of \citet{asplund2005};
\item covering the spread in chemical space of the low-$\alpha$ sequence with our synthetic model which takes into account the observational errors (see upper right panel of Fig. \ref{AGES_AFE_FEH});
\item reproducing the observed age distributions of the high-$\alpha$ and low-$\alpha$ sequences, and the MDF (see Figs. \ref{SAGE} and \ref{MDF}).
\end{itemize}

 In Fig. \ref{tmax} we note that  a  delay  of  $t_{max}$=1.3 Gyr at the beginning of the second gas accretion episode is similar to the one adopted in the  classical two-infall chemical evolution model presented by Chiappini et al. (2001) and Spitoni et al. (2009). Small differences between the two infall model presented in Fig. \ref{tmax}  with $t_{max}$=1.3 Gyr and the classical ones,   are due to the adopted model prescriptions: i.e in our model we did not considered a threshold in the SFR and the first infall time scale is shorter. Here we can conclude that the usual delay adopted by the classical two infall model does not properly apply to the new VSA18 stellar sample.
In Section \ref{s:t1t2} we will show how sensitive are the chemical evolution results to different choices for the infall time-scales $\tau_{D1}$ and $\tau_{D2}$. 

\begin{figure}
\begin{centering}
\includegraphics[scale=0.28]{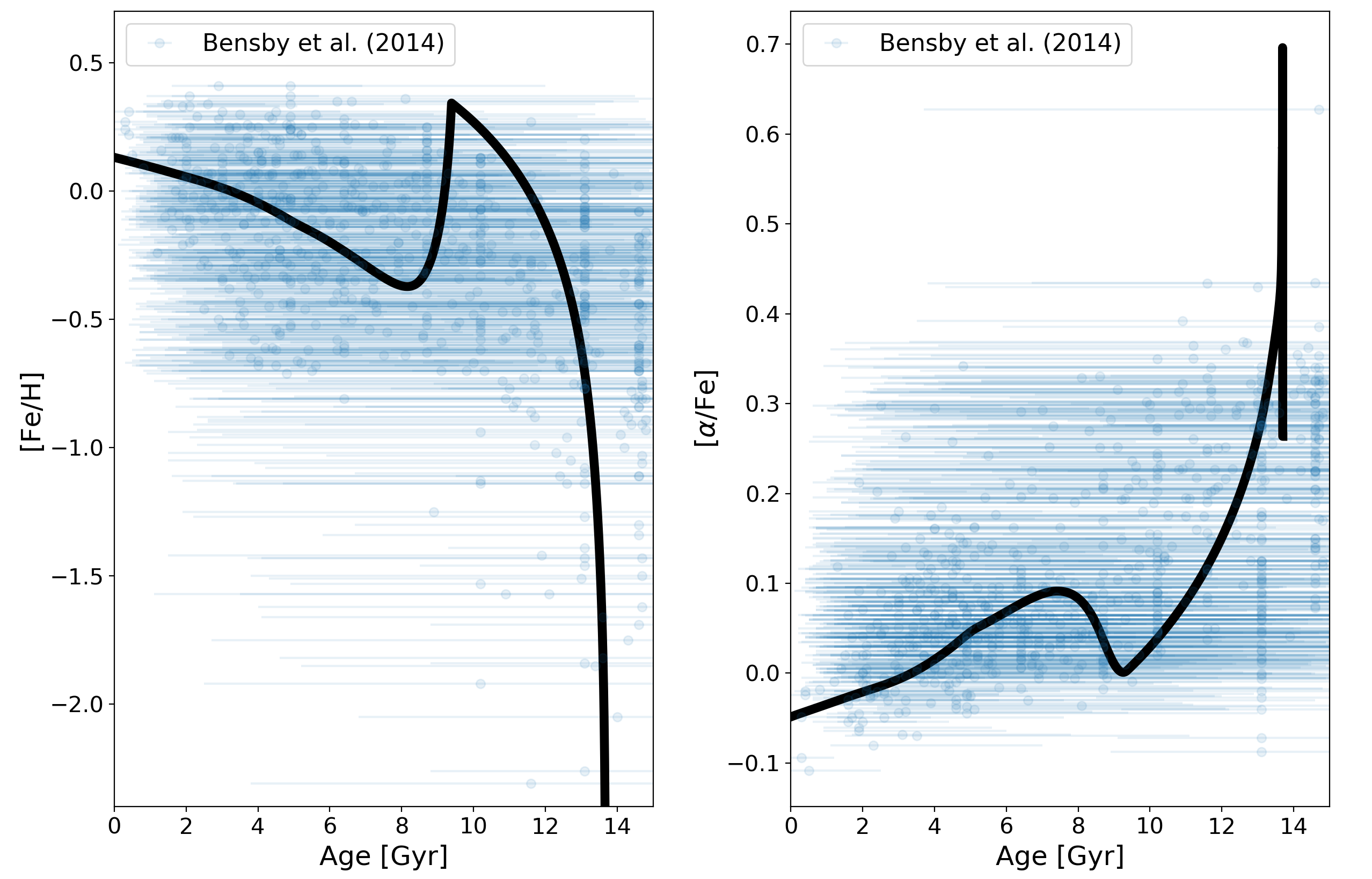}
  \caption{ Time evolution of [Fe/H] (left panel)  and [$\alpha$/Fe] (right panel)  predicted by our chemical evolution model (black solid lines)  compared with the  the observational values of the sample presented by \citet{bensby2014}.}
\label{bensby_age}
\end{centering}
\end{figure}

It is also worth noting that \citet{hayden2015} presented the [$\alpha$/Fe] vs. [Fe/H]  relation for APOGEE stars at different Galactocentric  distances. The correspondence between our results reported in Fig. \ref{AGES_AFE_FEH} for different stellar ages in the solar neighbourhood  and \citet{hayden2015} results (see their Fig. 4) is evident. In fact, according to  \citet{hayden2015} stars preferentially populate the low-$\alpha$ sequence in the [$\alpha$/Fe] vs. [Fe/H] in the outer Galactic regions, which is in agreement with our results for stars younger than 4 Gyr. In the light of our results and in presence of the stellar ages provided by asteroseismology, our analysis coupled with the observations presented by \citet{hayden2015} confirms  an inside-out formation scenario for the Galactic disc: outer Galactic regions have few old stars and show only recent episodes of star formation.
 Moreover, the \citet{hayden2015} data show that in the outer regions the locus of the low-$\alpha$ sequence shifts towards lower metallicity. 
This is can be well explained in by inside-out formation: external Galactic regions are formed on longer time scales, hence the chemical enrichment is weaker and less efficient than the inner Galactic regions, leading to a smaller metallicity.
As discussed in the Introduction, inside-out formation is  well motivated by the dissipative collapse scenario \citep{larson1976, cole2000}.
\begin{figure}
\begin{centering}
\includegraphics[width=\columnwidth]{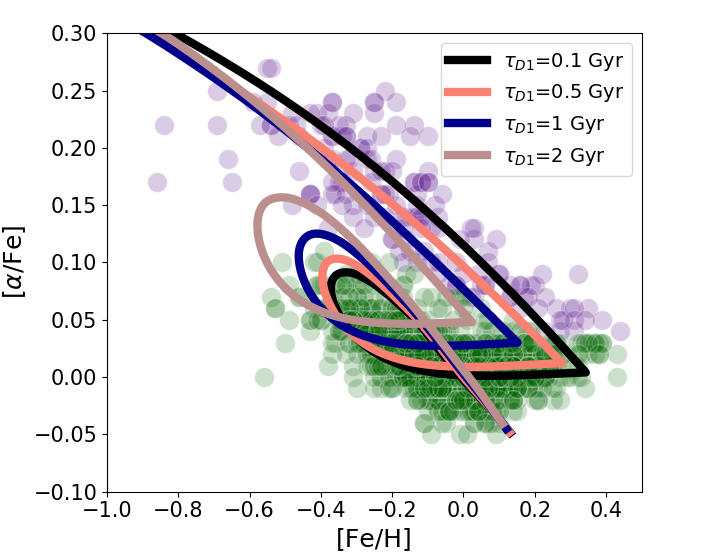}
\includegraphics[width=\columnwidth]{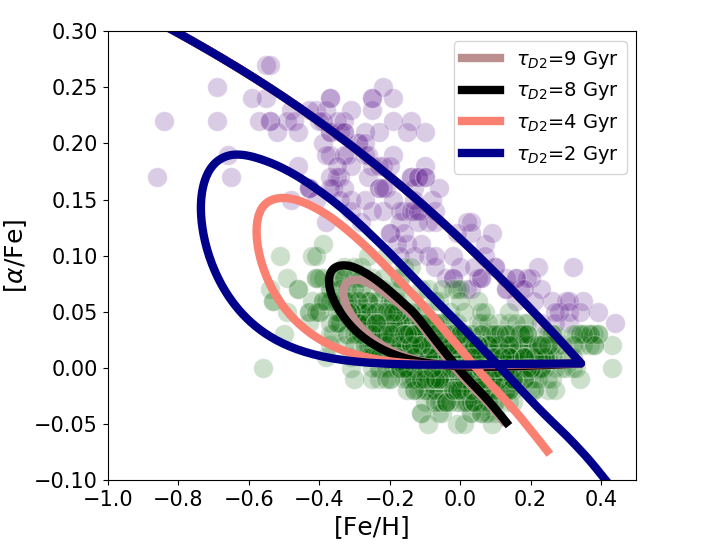}
\caption{Effects on the chemical evolution of the solar neighborhood in the [$\alpha$/Fe] versus [Fe/H] plane of different time-scales for primordial gas infall. {\it Upper panel}: time-scale variations for primordial gas accretion related to the high-$\alpha$ sequence ($\tau_{D1}$).
{\it Lower panel}: time-scale variations for  primordial gas accretion related to the low-$\alpha$ sequence ($\tau_{D2}$).
}
\label{T1}
\end{centering}
\end{figure}

\begin{figure*}
\begin{centering}
\includegraphics[scale=0.5]{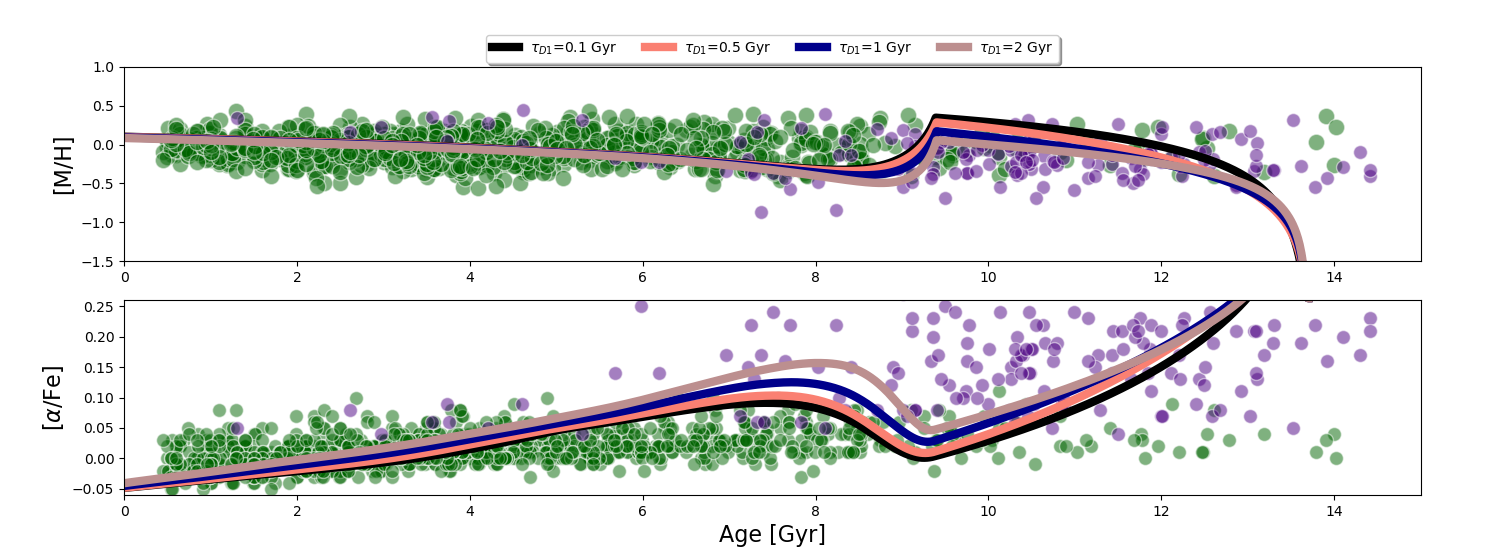}
\caption{Time evolution of [M/H] ({\it upper panel}) and [$\alpha$/Fe] ({\it lower panel}) shown for the VSA18 sample (same as Fig.~\ref{AFE_age_SD}) and our chemical evolution model with different time-scale values for the first infall of gas ($\tau_{D1}$)}
\label{T1_Age}
\end{centering}
\end{figure*}
\begin{figure*}
\begin{centering}
\includegraphics[scale=0.5]{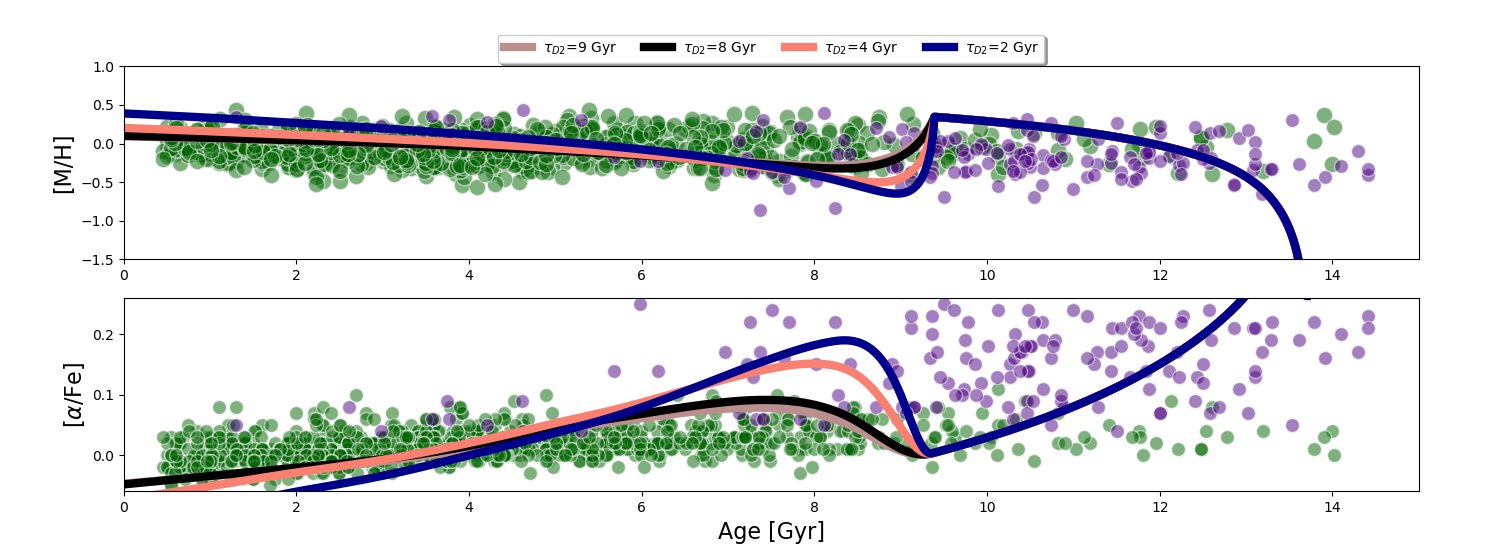}
\caption{
Same as Fig.~\ref{T1_Age} but for different time-scale values of the second infall of gas ($\tau_{D2}$).}
\label{T2_Age}
\end{centering}
\end{figure*}
In Fig. \ref{SAGE} we report our model results in terms of the stellar age distribution. In the upper panel the distribution of stars formed during  the whole Galactic history predicted by our ``two-infall'' chemical evolution models is drawn. The distinction between  old stars, whose distribution  peaks within the first Gyr of the Galactic time, and the young ones related to the second infall of gas is clearly shown. In order to compare our model results with the age distributions given by VSA18 we considered the errors introduced in
eqs. (\ref{Eqer1}) and (\ref{Eqer2}).

With the aim of comparing the observed high-$\alpha$ sequence with our
model, we want to consider only the stars formed up  to the time at which the
second infall start (i.e Galactic time $t < t_{max}$= 4.3
Gyr). 
For this purpose, we convolved our stellar age distribution with mock observational errors to create a new stellar age distribution.

Moreover, to be consistent with the data,
we considered only stars with [M/H] $>$ -1 dex. In the  middle panel of Fig. \ref{SAGE} we report this distribution along with the observed one. We note that the data are reasonably well reproduced, even if we predict more stars at the early times. 
The median of the observed high-$\alpha$  stellar ages distribution is $10.40^{+1.86}_{-2.71}$ Gyr, whereas the one predicted by our synthetic model  for  the  high-$\alpha$ sequence is  $10.53^{+2.23}_{-2.14}$ Gyr.
\begin{figure}
\begin{centering}
\includegraphics[scale=0.42]{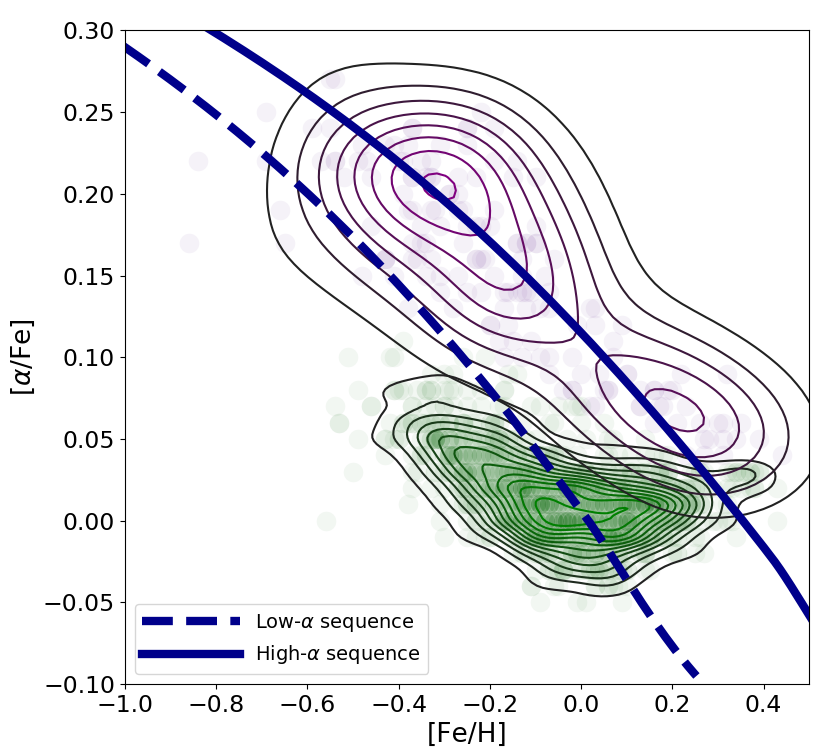}
\caption{Model results of the parallel approach for the abundance ratios  [$\alpha$/Fe] vs.  [Fe/H]. We report the evolution of the high-$\alpha$ (solid line) and the low-$\alpha$ (dashed line) model sequences. Observational data
are the same as in Fig.\ref{AFEFE1}.}
\label{par_AFE_FEH}
\end{centering}
\end{figure}

\begin{figure}
\begin{centering}
\includegraphics[scale=0.5]{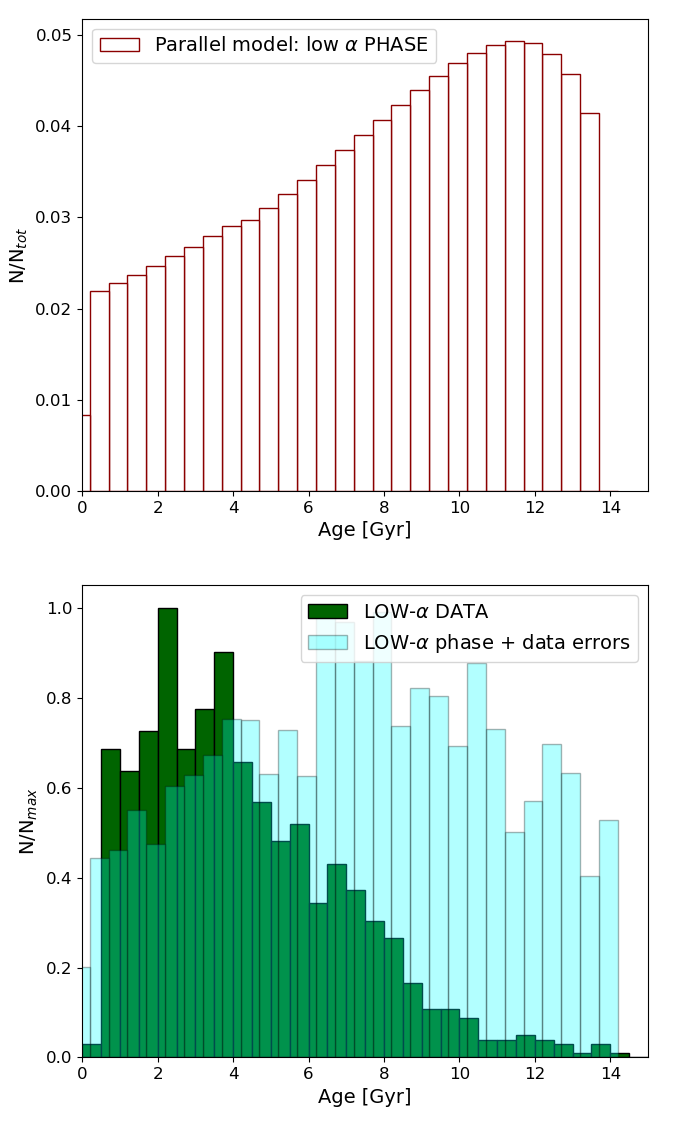}
\caption{{\it Upper panel}: stellar age distribution predicted by the parallel chemical evolution model for the  low-$\alpha$ phase.
{\it Lower panel}: age distribution for the low-$\alpha$ model sequence in which we have taken into account age and [M/H] errors (cyan histogram). Also plotted is the histogram of the age distribution for the low-$\alpha$ stars from VSA18 (green).}
\label{age_parallel}
\end{centering}
\end{figure}
\begin{figure}
\begin{centering}
\includegraphics[scale=0.7]{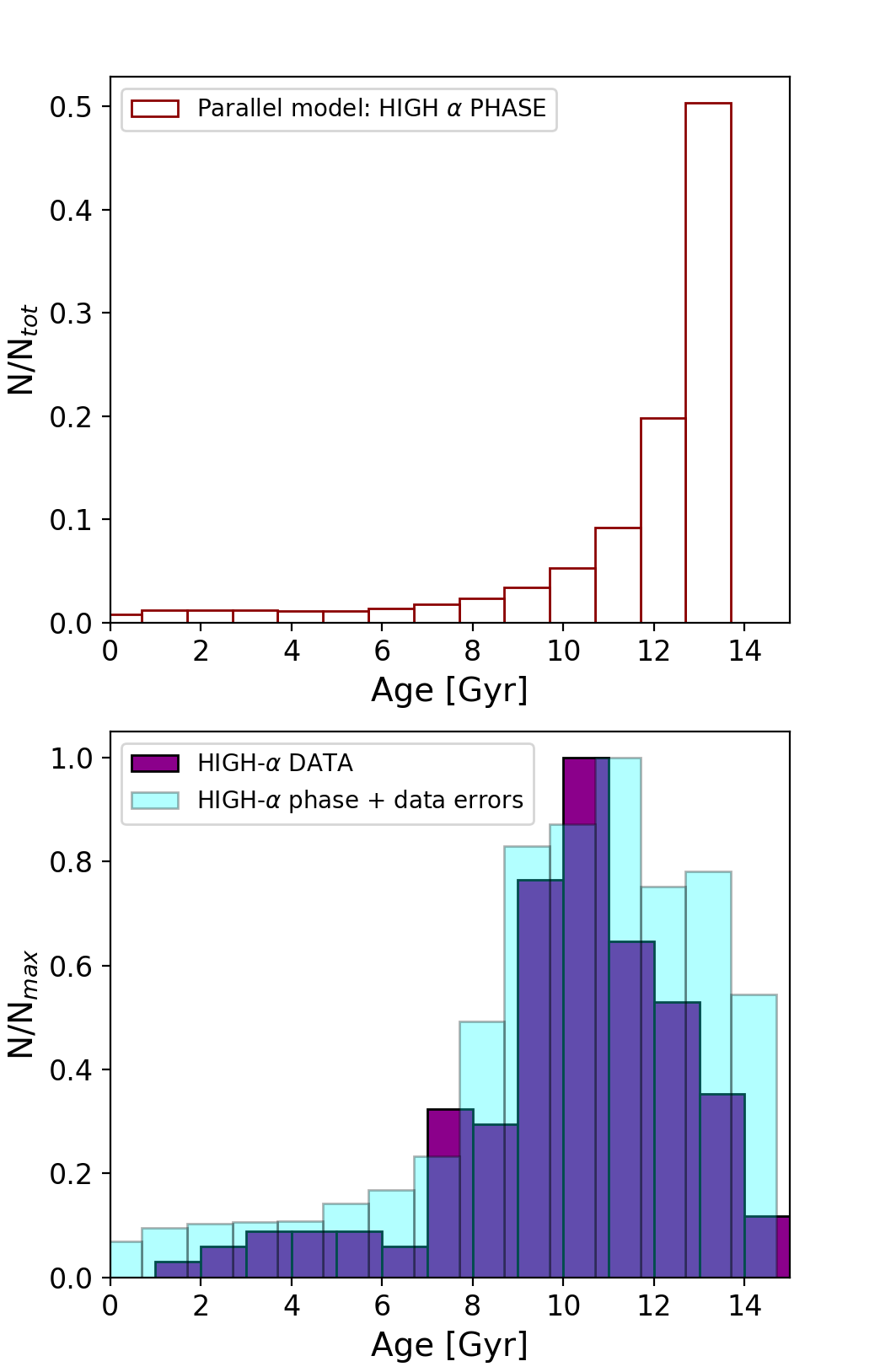}
  \caption{Same as Fig.~\ref{age_parallel} but for the high-$\alpha$ sequence. See text for details.}
\label{age_parallel_h}
\end{centering}
\end{figure}

In fact, in order to best
match the [$\alpha$/Fe] vs [Fe/H] abundance ratio we have to
consider a very fast evolution for the high-$\alpha$ stars. Anyway,
including the observational age error, our model predicts a spread in
the age distribution in agreement with the data one.

Adopting  the same method, the age distribution of the stars formed during the second infall of gas is compared with the APOKASC low-$\alpha$ sample in
the middle plot of Fig.~\ref{SAGE}. The general data trend is
reproduced, but some differences between model and observations can be
noticed also in this case: the observations show a distribution peaked
at slightly younger ages.
In fact, the median of the observed low-$\alpha$ stellar ages distribution is $3.76^{+3.32}_{-2.17}$ Gyr, whereas the one predicted by our synthetic model is  $4.31^{+3.16}_{-2.79}$ Gyr.

In Fig. \ref{MDF} the MDF of the
``two-infall'' model without taking into account any kind of observational errors
is compared with the whole data sample (low-$\alpha$ + high-$\alpha$
stars). The MDF is expressed in terms of the abundance ratio
[M/H] introduced in eq. (\ref{MH}).  It is evident that our model is
consistent with the data but predicts less stars at super-solar
metallicities compared to the data. The two distributions are
normalized to  the corresponding maximum number of stars for
each distribution.

In the same Figure we also show the curve related to the model distribution
convolved with a gaussian with a $\sigma$ fixed at the value of 0.13 dex, consistent with observational errors. The model line with the convolution (normalized at the total number of stars)  better reproduces the data as shown by the cyan line in Fig. \ref{MDF}. The model results related to the case in which we convolved our MDF with a gaussian with a $\sigma$  0.07 dex is also shown.
We recall that the average [M/H] error in APOGEE is $\sim$ 0.12 dex.   We show results for  2 values of sigma: one slightly  larger ($\sigma$ =0.13 dex) and another one smaller ($\sigma$= 0.07 dex) than the  average observational error at the aim to show how the distribution is affected by different $\sigma$  values.
In conclusion, this scenario is capable of reproducing almost all the observational proprieties. Contrary to \citet{Nidever:2014fj}, we do not need the
superposition of several populations with different enrichment histories, or
variable loading factor winds combined with different star formation
efficiencies in time. Our scenario is simpler and it is in agreement with the
stellar ages provided by asteroseismology.

 In Fig. \ref{bensby_age} we compare the age-metallicity and time evolution of [$\alpha$/Fe]  abundance ratio  of our updated two-infall model (without including errors) with the \citet{bensby2014} data. Our model is consistent with the data given the large uncertainty in the stellar age determinations. Comparing the age errors of Fig. \ref{bensby_age}  with the ones of the asteroseismology in  Fig. \ref{errors},  it is clear  that the asteroseismology opened a new era in the Galactic Asteroarchaelogy.
While our results are also consistent with the data by \citet{bergemann2014}, some discrepancy related to the high-$\alpha$ sequence emerges with the \citet{haywood2013} data. In that paper the high $\alpha$ stars show a tight relation in the  age vs [$\alpha$/Fe] in contrast with the finding by VSA18. 
Also our chemical evolution model predicts a tight correlation with a steeper slope,  however once we include  the observational errors this correlation disappears in the [$\alpha$/Fe] vs Age.  As underlined above,  regions with the higher density of high-$\alpha$ stars formed by our model (see lower panel of  the Fig. \ref{sdens})  overlap with the VSA18 data  (stars with ages between 8 and 14 Gyr and [$\alpha$/Fe] larger than 0.05 dex). 

The discrepancy between \citet{haywood2013} and VSA18 data could be due to the fact that \citet{haywood2013}  results are based on a subsample of \citet{Adibekyan2012} composed by only 363 stars with meaningful ages, corresponding to bright turn-off dwarfs (in \citet{Adibekyan2012} dwarfs are selected for exoplanet detection studies)  were no assessment has been made of how representative they are of the underlaying population.

\section{Testing different time-scales of accretion $\tau_{D1}$ and $\tau_{D2}$}\label{s:t1t2}
In this Section we test the impact on the chemical evolution of the solar neighbourhood of varying the time-scales of primordial infalling gas. Retaining  the same model prescriptions of the best model presented in Section~\ref{s:revised}, in upper panel of Fig. \ref{T1} we show the results of considering different values of
$\tau_{D1}$: 0.1, 0.5, 1, 2 Gyr. It is evident that as chemical enrichment is faster and more efficient (i.e., as $\tau_{D1}$ gets shorter):
\begin{itemize}
\item the high-$\alpha$ disc phase is shifted towards larger [Fe/H] values;
\item the system presents a lower [$\alpha$/Fe] value when the second infall starts (we recall that it takes place at $t_{max}$=4.3 Gyr). This is due to the fact that Type II SNe trace the SFR. If the SFR peaks at early time (e.g., $\tau_{D1}$=0.1), at $t_{max}$= 4.3 Gyr the iron produced by Type Ia SNe with a time delay will dominate the ISM pollution \citep{matteucci2009,bonaparte2013,vincenzo2017}. When the SFR is more extended in time (e.g., $\tau_{D1}$=2), a smaller [$\alpha$/Fe] abundance ratio is therefore expected.
\end{itemize}

 In Fig. \ref{T1_Age} we explore the age-metallicity (in terms of [M/H]) and age-$\alpha$ relations for models with different $\tau_{D1}$ values (same values as in upper panel of Fig. \ref{T1}). Models with shorter infall time-scales predict higher metallicities at the beginning of the second accretion phase ($t=t_{max}$=4.3 Gyr), while the model with the longest time-scale ($\tau_{D1}$= 2 Gyr) reaches the lowest metallicity after $t_{max}$ before the pollution by Type II SNe resumes.

In the lower panel of Fig.~\ref{T1_Age} we show the same models in terms of the predicted time evolution of the abundance ratio [$\alpha$/Fe]. At the moment of the start of the second infall ($t_{max}$=4.3 Gyr after the Big Bang), models with  shorter time-scales of gas accretion $\tau_{D1}$ have smaller SFR compared to models with longer time-scales. Therefore, the Type II SN
 contribution is smaller, whereas the Fe produced by Type Ia SNe with a  delay time distribution is important. Hence a smaller [$\alpha$/Fe] is expected as shown in Fig.~\ref{T1_Age}.

Always adopting the same model prescriptions of our best ``two-infall'' model (and $\tau_{D1}$=0.1), in the lower panel of Fig.\ref{T1} we show the results when we vary the second infall time-scale $\tau_{D2}$ assuming the following values: $\tau_{D2}=2, 4, 8,$~and 9 Gyr. It is evident that the size of the ``loop'' is strongly dependent on the time-scale of gas accretion $\tau_{D2}$. In fact, at the beginning of the second accretion event, the infall rate of pristine gas is higher for smaller $\tau_{D2}$ and the therefore the dilution effect (longer horizontal excursion towards lower [Fe/H] values) is more evident. Consequently, a more extended loop in the [$\alpha$/Fe] vs. [Fe/H] relation appears as a result of the  larger increase in [$\alpha$/Fe] produced by SNe type II for smaller $\tau_{D2}$.

In Fig.~\ref{T2_Age} we show the age and $\alpha$ abundances evolution compared to our models calculated with different $\tau_{D2}$ values (same models as those shown in the lower panel of Fig. \ref{T1}). In systems with shorter time-scales of accretion $\tau_{D2}$, the rate of gas accretion during the second infall is large at Galactic times $t \simeq t_{max}$, hence the dilution works efficiently. In fact, in the upper panel of Fig. \ref{T2_Age} we see that the model with $\tau_{D2}$= 2 Gyr presents the deepest drop in the age vs. [M/H] relation.

\begin{figure*}
\begin{centering}
\includegraphics[scale=0.5]{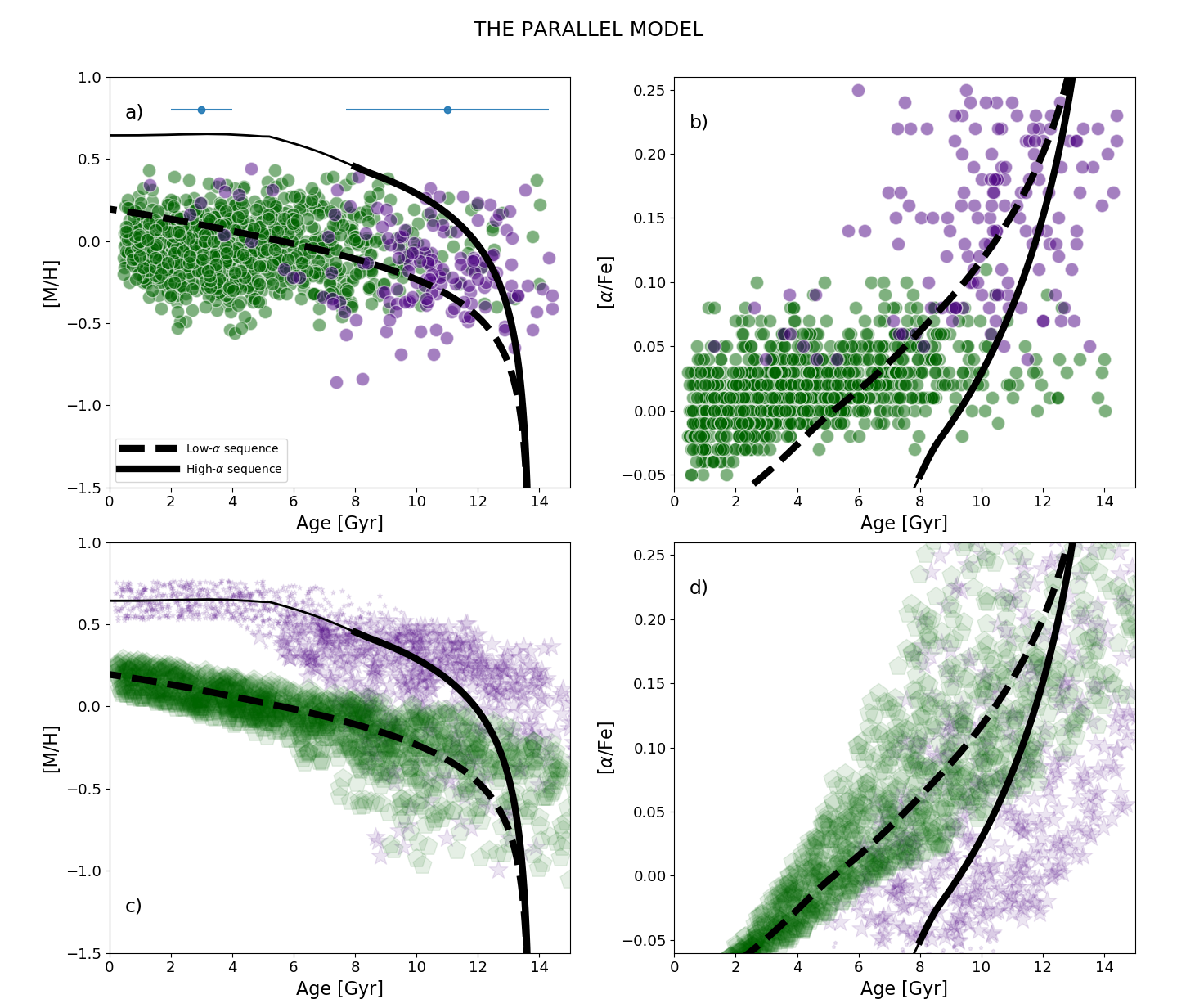}
 \caption{  Panels a) and b): same as Fig.~\ref{AFE_age_SD} for the parallel model results. 
The dashed black line depicts the low-$\alpha$ model phase, whereas the black solid line stands for the high-$\alpha$ phase. In the latter case, the thin line shows the period of the evolution when the number of stars born in age bins of 1 Gyr is smaller than 2\% of total number of stars formed.  Panels c) and d): chemical evolution model results where we have taken into account estimated age and [M/H] errors. Analogously to the thin solid line in the upper two panels, smaller purple  starred symbols  are associated to stars with ages lower than 8 Gyr (before the inclusion of age errors). See text for details.}
\label{parallel4}
\end{centering}
\end{figure*}
\begin{figure}
\begin{centering}
\includegraphics[scale=0.43]{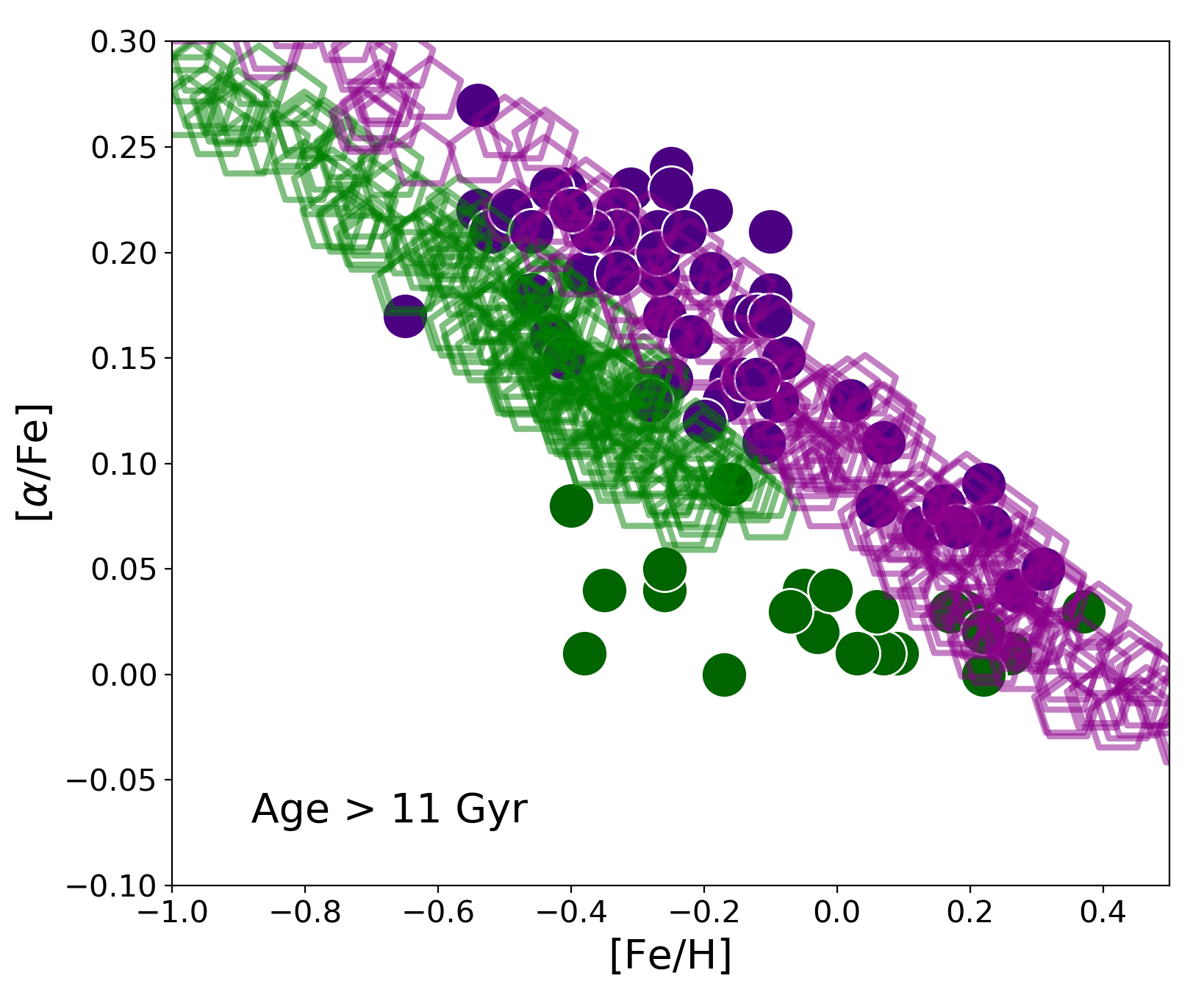}
  \caption{[$\alpha$/Fe]  abundance ratios as a
  function of [Fe/H]
predicted by means of the parallel chemical evolution model including age and  [M/H]
errors  for the high-$\alpha$ and low-$\alpha$ model phases (empty magenta 
and green pentagons, respectively). 
 Both
 models and data are related to stars older than 11 Gyr. }
\label{parallel11}
\end{centering}
\end{figure}

In the lower panel of Fig. \ref{T2_Age},  the model with the highest
bump in the [$\alpha$/Fe]  abundance ratio is the one  with the shortest timescale
of accretion  $\tau_{D2}$. As briefly mentioned before, the star formation activity is tightly connected with the rate of Type II SNe, and hence to the $\alpha$-element
production. On the other hand, Fe production needs a certain time delay and this is the reason why the model curve with $\tau_{D2}$=2 Gyr shows the steepest
increase in the [$\alpha$/Fe]  soon after the beginning of the second infall.

\section{Model results for the ``parallel'' formation scenario}
Following the scenario proposed by \citet{grisoni2017} we show here the results for the ``parallel'' model. We recall that in this scenario the high-$\alpha$ and low-$\alpha$ sequences evolve i) independently ii) and coevally.

In Fig. \ref{par_AFE_FEH} we show the results of chemical evolution model in the the [$\alpha$/Fe] vs. [Fe/H] plane. Following Eqs.~\ref{par1} and~\ref{par2}, the time-scales of gas accretion are $\tau_{T}$= 0.1 Gyr and $\tau_{D}$= 7 Gyr, assuming the same values adopted by \citet{grisoni2017}.

The SFEs for the high-$\alpha$ and low-$\alpha$ sequences are fixed at
the values of $\nu$ =1.3 Gyr$^{-1}$ and $\nu$ =0.7 Gyr$^{-1}$, respectively, and the current total surface mass densities are the same as those from our revised "two-infall" model. An overall good level of agreement is obtained between the data and model: the high-$\alpha$ sequence is reproduced by a  fast and efficient evolution, whereas the low-$\alpha$ stars are better characterized by a slower evolution \citep[as shown by][]{grisoni2017}.

However, this scenario presents several flaws in the light of the new key information as provided by asteroseismology: stellar ages. In fact, in Fig. \ref{age_parallel} we present the age distribution  of the low-$\alpha$ sequence predicted by our chemical evolution model which also takes into account the observational errors described by Eqs.~\ref{Eqer1} and~\ref{Eqer2}.

We notice that the age distribution of the stars formed in the low-$\alpha$ model is very different from the one presented by VSA18: too many stars in the parallel approach are created at early times, whereas the observed distribution shows that  the majority of them has ages between between 2 and 4 Gyr. We recall that  the median of the observed low-$\alpha$ age distribution is located at  $3.76^{+3.32}_{-2.17}$ Gyr, whereas the one predicted by our synthetic  parallel model is  $7.06^{+4.13}_{-3.76}$ Gyr.

On the other hand, in Fig.~\ref{age_parallel_h} we show that the age distribution predicted by  the high-$\alpha$ sequence of the parallel approach is in agreement with the observational data.
The median of the observed high-$\alpha$  age distribution is $10.40^{+1.86}_{-2.71}$ Gyr, whereas the one predicted by our  high-$\alpha$ synthetic parallel model  sequence is  $10.12^{+2.09}_{-2.94}$ Gyr.

These results are further analysed in Fig.~\ref{parallel4}, where we show the evolution of metallicity and [$\alpha$/Fe] for the parallel model. The thin  line in the high-$\alpha$ sequence represents the chemical evolution phase in which the number of stars formed in age bins of 1~Gyr is smaller than 2\% of total number of stars created thorough the whole Galactic life (the age distribution in upper panel of Fig. \ref{age_parallel_h} shows that it is true for Galactic ages smaller than $\sim$ 8 Gyr).

We notice that the high-$\alpha$ sequence is characterized by a fast and efficient chemical enrichment, with high values of [M/H] compared to the low-$\alpha$ model. In the two lower panels of Fig. \ref{parallel4} the time evolution of metallicity [M/H] and [$\alpha$/Fe] ratios including the errors described in eqs. (\ref{Eqer1}) and (\ref{Eqer2}) are reported for the parallel model. The metallicity predicted by the model for the high-$\alpha$ sequence at young ages is much higher than the observed metallicity distribution of stars. However, given that the fraction of young stars formed in the high-$\alpha$ sequence is negligible compared to the old ones (c.f., Fig. \ref{age_parallel_h}), these stars are unlikely to be observed.

Looking at the evolution of [$\alpha$/Fe] predicted by the parallel model, we see significant differences compared to the observed distribution obtained from asteroseismology. In fact, at early times the low-$\alpha$ model predicts that  the majority of the stars should have high [$\alpha$/Fe] values. 
Moreover, the low-$\alpha$ sequence predicts at young ages  stars with much lower [$\alpha$/Fe] values than the ones observed.

A more evident tension between the observational data and the models
is present in [$\alpha$/Fe] vs. [Fe/H] plane for stars older than 11 Gyr, as shown in Fig. \ref{parallel11}. After including the uncertainties in age and metallicity, the model cannot reproduce the population of stars at $-0.5<$[Fe/H]$<0.25$ and [$\alpha$/Fe]$\leq$0.05. Even if this population of old low-$\alpha$ stars are missclassified due to large age uncertainties, the revised "two-infall" model was capable of predicting the existence of such   objects once appropriate errors were included in the model calculations. 
In the parallel model there is the lack of an horizontal stripe in the [$\alpha$/Fe] vs. [Fe/H] plane, which characterizes the observational data at different Galactic ages (see data in Fig. \ref{AGES_AFE_FEH}).

We conclude that the purely parallel approach fails to reproduce the data in the solar neighborhood if we take into account the new dimension provided by the asteroseismology, i.e. the stellar age.
\section{Conclusions}\label{conc}
We have studied in detail chemical evolution models in the solar annulus with the aim of reproducing the new observational data by \citet{2018MNRAS.475.5487S}, concerning both chemical abundance ratios and precise stellar ages as provided by asteroseismology. Our main conclusions can be summarized as follows:
\begin{itemize}
\item Our revised ``two-infall'' model in the solar neighborhood well reproduces the observational stellar properties of both the high-$\alpha$ and low-$\alpha$ sequences;
\item The APOGEE data is consistent with the presence of a delayed second infall of gas, which in our model creates a loop in the [$\alpha$/Fe] vs  [Fe/H] plane that corresponds to low-$\alpha$ stars;
\item With the inclusion in our model of the observational age and metallicity uncertainties we nicely reproduce i) the spread in the age-metallicity relation, and ii) the time evolution of the [$\alpha$/Fe] abundance ratio. Moreover, the observed stars older than 11 Gyr seem to confirm our Astroarcheology scenario. In fact, these stars keep the signature of a second infall of gas delayed by $\sim$4.3 Gyr with respect to the first episode and the successive dilution effect in the [$\alpha$/Fe] vs. [Fe/H] plane;
\item Our revised ``two-infall'' model results are in agreement with the observed age distribution of the stars of the high-$\alpha$  and low-$\alpha$ sequences,  and with the observed metallicity distribution function;
\item We showed that the ``parallel'' model, in which the high-$\alpha$ and low-$\alpha$ sequences form coevally but independently with different time-scales of accretion, is not able to reproduce the constraints given by the stellar ages. In fact, the low-$\alpha$ sequence model  cannot reproduce the location in the [$\alpha$/Fe] vs. [Fe/H] plane of stars older than 11 Gyr even when observational uncertainties are taken into account. Moreover, the low-$\alpha$ sequence predicts, at young ages,  stars with much lower [$\alpha$/Fe] values than the ones observed.
\end{itemize}

 By means of chemical evolution models we provide constraints on the accretion history of the Milky Way in the light of new observational data. We showed that the two infall model is still a valid one but we pointed out the importance of a consistent delay in the second accretion ($t_{max}$ = 4.3 Gyr) to properly reproduce the properties of the low-$\alpha$ stars.  The presence of an horizontal sequence in the [$\alpha$/Fe]-[Fe/H] plane was predicted by Calura \& Menci (2009) by means of chemical evolution models in a cosmological framework. 
A new assessment of such a feature by means of up-to-date galaxy formation models is to be considered for future work.   Our results are consistent with very high resolution cosmological zoom-in AURIGA simulations for Milky Way sized haloes presented by \citet{grand2018}. They found that a bimodal distribution in the [$\alpha$/Fe] vs. [Fe/H] plane is due to the presence  of a  temporarily lowered gas accretion rate. In our ``two infall'' model  a lowering of the gas accretion is mimicked by a consistent delay in the second infall of gas.

Nevertheless, the scenario presented in this work is not complete. In fact, we ignored in our analysis the Y$\alpha$R stars, which are believed to be  originated by an interacting binary stellar system or stellar migration. We did not discuss the effects of stellar migration, which plays an important role, but focused on providing a simple formation scenario yet able to reproduce the new tight
constraints provided by the asteroseismic stellar ages. Further analysis including stellar kinematics will be the subject of an upcoming publication.
\section*{Acknowledgement}
 We thank the anonymous referee for various suggestions that improved the paper.  E. Spitoni thanks  K. Verma for helpful discussions.  E. Spitoni and V. Silva Aguirre acknowledge support from the Independent Research Fund Denmark (Research grant 7027-00096B). V. Silva Aguirre acknowledges support from VILLUM FONDEN (Research Grant 10118). F. Matteucci  acknowledges research funds from the University of Trieste (FRA2016). F. Calura acknowledges funding from the INAF PRIN-SKA 2017 program 1.05.01.88.04.

\bibliographystyle{aa} % style aa.bst
\bibliography{disk}

\end{document}